\documentclass{elsarticle}
\usepackage[margin=2.5cm]{geometry}

\usepackage{multirow}
\usepackage{lineno}
\usepackage[dvipsnames,table]{xcolor}
\usepackage[utf8]{inputenc}
\usepackage{lineno}
\usepackage[hidelinks]{hyperref}
\usepackage{gensymb}
\usepackage{textcomp}
\usepackage{amsmath}
\usepackage{graphicx}
\usepackage{amsfonts}
\usepackage{booktabs}
\usepackage{subcaption}
\usepackage{soul}
\usepackage{tabularx}
\usepackage{natbib}
\usepackage{color}
\usepackage{comment}
\usepackage{algorithm,algorithmic}
\usepackage{esvect}

\begin{document}

\begin{frontmatter}

\title{Development and Evaluation of an Online Home Energy Management Strategy for Load Coordination in Smart Homes with Renewable Energy Sources}

%% Group authors per affiliation:

%% or include affiliations in footnotes:
\author[mymainaddress]{Xiaoling Chen}

\author[mymainaddress]{Cory Miller}

\author[mymainaddress]{Mithun Goutham}

\author[mysecondaryaddress]{Prasad Dev Hanumalagutti}

\author[mysecondaryaddress]{Rachel Blaser}

\author[mymainaddress]{Stephanie Stockar\corref{mycorrespondingauthor}}

\cortext[mycorrespondingauthor]{Corresponding author}
\ead{stockar.1@osu.edu}

\address[mymainaddress]{Center for Automotive Research, The Ohio State University, Columbus, OH 43210, USA}
\address[mysecondaryaddress]{The Ford Motor Company, Dearborn, MI 48126, USA}

\begin{abstract}
In this paper, a real time implementable load coordination strategy is developed for the optimization of electric demands in a smart home. The strategy minimizes the electricity cost to the home owner, while limiting the disruptions associated with the deferring of flexible power loads. A multi-objective nonlinear mixed integer programming is formulated as a sequential model predictive control, which is then solved using genetic algorithm.
The load shifting benefits obtained by deploying an advanced coordination strategy are compared against a baseline controller for various home characteristics, such as location, size and equipment. The simulation study shows that the deployment of the smart home energy management strategy achieves approximately 5\% reduction in grid cost compared to a baseline strategy. This is achieved by deferring approximately 50\% of the flexible loads, which is possible due to the use of the stationary energy storage.
\end{abstract}

\begin{keyword}
Home Energy Management Strategy,
Load Coordination, 
Renewable Energy, 
%Energy Storage,
Decentralized Model Predictive Control
%Decentralized Control,
%Sequential Scheme, 
%Genetic Algorithm
\end{keyword}

\end{frontmatter}

\section{Introduction}

The global energy demand is predicted to have a remarkable increase for the next two decades \cite{conti2016international}, and the residential energy usage constitutes almost 40\% of the overall electricity consumption \cite{shewale2022survey}.
The growing power demand makes it challenging for grid reliability, which is ensured by keeping a balance between grid power generation and residential energy demand \cite{sarker2021progress}.
Moreover, the power demand of the residential sector is characterized by fluctuating and unpredictable demands due in part to the growing presence of local renewable energy sources \cite{liang2016emerging}, and increasing adoption of electric vehicles \cite{shao2009challenges}.
Among the smart grid infrastructure, demand-side management (DSM) is one of the most common strategies adopted to help curb grid reliability issues  \cite{shewale2020overview,energy_gov_2021}. 
Specifically,  time of use (TOU) and real time pricing schemes \cite{ATUS} have been used for shaping the power demand and increasing the residential demand predictability.  These pricing schemes encourage home owners with cost incentives to use less electricity during high demand times to flatten the peak power curve. 
In this context, a home energy management (HEM) strategy automates the scheduling by controlling flexible loads, such as laundry and dishwasher operation, heating ventilation and air conditioning, renewable energy sources and energy storage systems \cite{costanzo2011peak}.
%Offline optimization approach requires exact knowledge of future inputs into the HEM system, such as the ambient temperature and load demands to determine the optimal HEM strategy \cite{paterakis2015optimal}.
Online HEM approaches do not require a-priori knowledge of future power demands, such as the ambient temperature or appliance oprations, and are therefore suited for practical deployment. 
In this scenario, Model Predictive Control (MPC) is often utilized due to its ability to account for the dynamics of the system, input and state constraints, uncertainties in the system and ability to minimize a performance metric \cite{beaudin2015home}.
In the area of smart home technology, MPC has been proven to achieve satisfactory performance in heating \cite{karlsson2011application}, cooling \cite{ma2009model} and ventilation \cite{yuan2006multiple}. 
However, due to the size of the problem and nonlinear dynamics of a smart home, using exact approaches for optimal load coordination is often computationally intractable \cite{beaudin2015home}. 
For this reason, many online approaches for load coordination rely on heuristic and meta-heuristic algorithms \cite{hassan2005comparison,rahmani2016energy,hu2018price,huang2019hybrid,jiang2019household}. 

For example, \cite{zhao2013optimal} presents an MPC with genetic algorithm (GA) to reduce the electricity cost and peak to average ratio by optimizing HVAC, laundry, dishwasher and non-deferrable loads. The uncertain nature of integrated renewable energy sources is considered in \cite{li2018real,javaid2018energy,arabali2012genetic}, where the load scheduler optimizes the charge-discharge strategy of a coupled energy storage system using a GA.
More comprehensive studies on load coordination using meta-heuristic approaches are presented, for example in \cite{hu2018price} where genetic algorithm was applied for the selection of temperature set points of the residential heating, ventilation, and air conditioning (HVAC) unit to reduce the electricity bills together with peak load. The controlled appliances are extended in \cite{jiang2019household} to include laundry and dishwasher, and their delay in activation. However, all these strategies were implemented offline, where all user requests were known a-priori. 
In summary, current HEM strategies either focus on specific appliances and can be implemented online or consider multiple loads and renewable sources, but are only applicable offline. Hence, there is a lack of a comprehensive online implementable HEM strategy that coordinates all flexible loads.

Moreover, many studies analyzing different HEM strategy performance use case specific evaluation metrics \cite{yildiz2017recent,veras2018multi,beaudin2015home,brahman2015optimal,thomas2018optimal}. The lack of generalizability of the performance metrics, makes it impossible to perform a comprehensive evaluation of benefits from home energy management strategy under different conditions, such as house location and seasonality. To evaluate the controller performance in a variety of scenarios, a more general performance metric, namely the deferral efficiency, is first introduced in this paper.

This paper presents a novel online scheduling strategy based on MPC that is used to realize the DSM by managing load coordination in the presence of both local renewable generation and an energy storage. Specifically, the smart home is assumed to include an HVAC system, plug-in and battery electric vehicle (xEV) charging, laundry and dishwasher, solar panels, and a stationary energy storage system.
The online implementable strategy is developed using a hierarchical structure, where the optimization problems are solved using a meta-heuristic approach. This structure achieves faster-than-real-time computation time and near-optimal solutions also under large variations in the home features and equipment. The performance of the controller is evaluated by conducting year-long simulations with significant variations in the household characteristics.

Results show that, the deployment of the smart HEM strategy achieves at least 5\% reduction in grid cost compared to the baseline strategy. The absolute savings are largely affected by the specific pricing scheme considered. In fact, the proposed HEM strategy is capable of deferring approximately 50\% of the flexible loads. This is accomplished through a more frequent utilization of energy storage to benefit from low grid electricity cost, and a more effective utilization of renewable sources.

% Need to be more clear on the novelty and contributions
% 1. The MPC sequential scheme in combination with  GA allows to compensate for uncertainties while achieving fast computation time. This means that the controller can be deployed in real-word applications, but also that we can perform a large scale simulation campaign within a reasonable amount of time.
% 2. The modularity of the controller allows for achieving near-optimal solution for a variety of simulation scenario. This controller is not ad-hoc and hence it allows to perform large scale simulation studies with significant variations in the parameters (location)
% 3. Because this controller works well on a variety of scenarios, a more general definition of performance metric should be introduced --> deferral efficiency

The remainder of the paper is structured as follows. First, the smart home plant model is described and a baseline control is presented. 
Then, the development of the new HEM strategy is discussed, together with the application of the sequential scheme and GA integration. 
Next, a comprehensive simulation campaign is conducted to investigate the performance of the real time controller under different plant model settings. Finally, the results of the simulation study are summarized and discussed with future steps.

\section{Home Plant Model} \label{sec:ModelDescription}

\begin{figure}
\centering
%trim={<left> <lower> <right> <upper>}
    \includegraphics[trim={7.8cm 5cm 5cm 0cm},clip, width= 0.5\columnwidth]{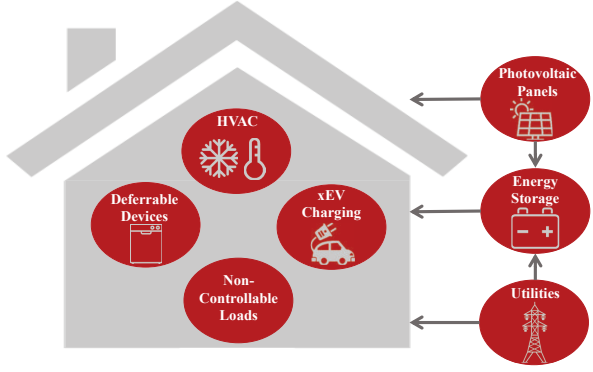}
    \caption{Home Plant Model}
    \label{fig:HomeModel}
\end{figure}

The residential power consumption is determined by predicting controllable household appliance loads, such as HVAC, smart devices, and xEV charging; together with base user activities, such as cooking, cleaning, leisure \cite{maiorino2015modeling,goutham2020machine}. The household power sources are the electric grid and local renewable generation from solar panels.  or the discharge. The household model and components are shown in Figure \ref{fig:HomeModel}. 
The associated household power balance is:
\begin{equation} \label{eq:PowerBalanceModel}
P_{Grid}+P_{Solar}+P_{ES}=P_{HVAC}+P_{xEV}+P_{D}+P_{ND}
\end{equation}
where $P_{Grid}$ is the grid power, $P_{Solar}$ is the power generated locally, $P_{ES}$ is battery power, $P_{HVAC}$ is the power for the operation of HVAC, $P_{xEV}$ is the power to charge the vehicle, $P_{D}$ is the power load from controllable appliances, and $P_{ND}$ is the non-deferrable power. It is worth noting that in the above formulation of power balance, a positive energy storage power $P_{ES}$  corresponds to discharging the battery.

\subsection{Photovoltaic System} \label{subsection:Photovoltaic}
The power generated by the photovoltaic panels is calculated using:
\begin{equation} \label{eq:solarEnergyUseful}
    P_{Solar}(t)= G(t) \cdot A_{PV} \cdot \eta_{PV} 
\end{equation}
where $G(t)$ is the useful solar irradiation power density, $A_{PV}$ is the total area of the solar panels, and $\eta_{PV}$ is the efficiency of the solar panel, which is assumed to be a function of the ambient and panel temperature,  \cite{cristaldi2012simple}. The useful solar irradiation power density, $G(t)$ is mapped based on location, season, and panel tilt \cite{goutham2020machine}. 

\subsection{Residential Energy Storage} \label{subsection:ResidentialES}
The power associated with the stationary energy storage is:
\begin{equation} \label{eq:ESPower}
    P_{ES}(t) = 
    \begin{cases}
        V_{ES}(t) \cdot I_{ES}(t) \cdot \eta_{ES} & I_{ES}(t) > 0\\
        V_{ES}(t) \cdot I_{ES}(t) / \eta_{ES} & I_{ES}(t) < 0\\
    \end{cases}
\end{equation}
where $\eta_{ES}$ is the constant battery charging and discharging efficiency, $V_{ES}$ is the pack voltage, and $I_{ES}$ is the pack current. 

The energy storage voltage is:
\begin{equation} \label{eq:ESV}
    V_{ES}(t)=V_{ES,cell}(t) \cdot N_{ES,series}
\end{equation}
where $N_{ES,s}$ is the number of cells in series, $V_{ES,cell}$ is the cell voltage determined using a 0\textsuperscript{th}-order equivalent circuit model \cite{lam2011practical}: 
\begin{equation} \label{eq:ESVoltage}
    \begin{split}
         V_{ES,cell}(t)=V_{OC}&(SOC_{ES},T_{ES})-R_0(SOC_{ES},T_{ES}) \cdot I_{ES,cell}(t)
    \end{split}
\end{equation}
where $I_{ES,cell}$ is the current applied to individual cells, $R_0$ is the internal resistance of each cell which is a function of state of charge (SOC) and temperature of the pack, and $V_{OC}$ is the open-circuit voltage, which is function of SOC and pack temperature. Internal resistance and open circuit voltage are mapped using \cite{lam2011practical}.
%Due to the low currents applied to residential energy storage systems, the heat generation of the batteries is assumed to be negligible and a thermal model is not included for the energy storage system. Instead 
The pack temperature is assumed to be either the same as ambient temperature if stored outside, or the same as the house temperature if stored inside. %An extension of this model that includes the thermal behavior of the energy storage system was presented in \cite{miller2022dc}. 
The cell current is calculated as:
\begin{equation} \label{eq:ESC}
    I_{ES,cell}(t)= \frac{I_{ES}(t)}{N_{ES,p}}
\end{equation}
Where $N_{ES,p}$ is the number of cells in the pack in parallel. Finally, the SOC is determined us
\begin{equation} \label{eq:ESSOC}
    \frac{dSOC_{ES}(t)}{dt}=-\frac{I_{ES,cell}(t)}{Q_{cell}}
\end{equation}
where $Q_{Cell}$ is the nominal cell capacity.In this paper, it is assumed that the ES starts empty, hence the initial SOC is 20\%. 
The number of cells in series and parallel are selected to meet the desired pack capacity $Q_{ES,des}$ and voltage $V_{ES,des}$:
\begin{equation} \label{eq:ESSizingParallel}
    Cells_{ES,parallel}= \frac{Q_{ES,des}}{Q_{cell}}
\end{equation}
\begin{equation} \label{eq:ESSizingSeries} 
    Cells_{ES,series}=\frac{V_{ES,des}}{V_{cell}}
\end{equation}
In this paper, the nominal ES size is 13.5kWh, 50V. Depending on the size of the house, multiple energy storage units are added \cite{goutham2020machine}.

\subsection{HVAC} \label{subsection:HVAC}
The power of the HVAC is calculated for each season:
\begin{equation} \label{eq:HVACEnergyConsumptionTotal}
    P_{HVAC}(t)=
    \begin{cases}
            P_{heat}+P_{fan} & T_{\infty} \leq T_{ref}\\
            P_{cool}+P_{fan} & T_{\infty} > T_{ref}\\
        \end{cases}
\end{equation}
where $T_{\infty}$ is the environment temperature, and $T_{ref}$ is a threshold temperature that determines whether the HVAC operates in heating or cooling mode \cite{maiorino2015modeling}. The  power consumption is given by:
\begin{eqnarray} \label{eq:HVACEnergyConsumptionHeat}
    P_{heat}&=\frac{\dot{m}_{HVAC} \cdot \pi(t) \cdot c_p \cdot \Delta T}{COP(\Delta T)}\\
    \Delta T&=T_{HVAC}-T_{\infty}
\end{eqnarray}
and
\begin{eqnarray} \label{eq:HVACEnergyConsumptionCool}
    P_{cool}&=\frac{\dot{m}_{HVAC} \cdot \pi(t) \cdot c_p \cdot \Delta T}{SHR \cdot COP(\Delta T)}\\
    \Delta T&=T_{\infty}-T_{HVAC}
\end{eqnarray}
where $\dot{m}_{HVAC}$ is the mass flow rate, $\pi$ is the fan command, $c_p$ is the air specific heat at constant pressure, $\Delta T$ is the temperature difference between the HVAC supply air temperature $T_{HVAC}$ and ambient temperature, COP is the Coefficient of Performance, and SHR is the Sensible Heat Ratio \cite{maiorino2015modeling}. 
For heating mode, the fan command $\pi$ can only be on or off, while in cooling mode, the fan command $\pi$ can be on, off, or half load. The fan regulation power is:
\begin{equation} \label{eq:HVACEnergyConsumptionFan}
    P_{fan}=\frac{\dot{m}_{HVAC} \cdot \Delta P}{\eta_{fan} \cdot \rho_{air}}
\end{equation}
where $\Delta P$ is the static pressure drop, $\eta_{fan}$ is the constant efficiency of the fan, and $\rho_{air}$ is the air density.

The household temperature dynamics is
\begin{equation} \label{eq:thermal}
\begin{split}
      m_ac_v\frac{d}{dt}T_a(t)=\pi(t)\cdot \dot{m}_{HVAC}\cdot c_p[&T_{HVAC}-T_a(t)]
      -\frac{T_a(t)-T_\infty}{R_{tot}}  
\end{split}
\end{equation}
where $m_a$ is the  mass of the air inside the house, $c_v$ is the air specific heat capacity at constant volume, $T_a$ is the air temperature inside the house, and $R_{tot}$ is the equivalent thermal resistance \cite{muratori2014dynamic}.

\subsection{Plug-in Electric and Electric Vehicle Charging} \label{subsection:xEV}
The power required to charge the vehicle is
\begin{equation} \label{eq:xEVPower}
    P_{xEV}(t) = \frac{V_{xEV}(t) \cdot I_{xEV}(t) }{ \eta_{xEV}}
\end{equation}
Where $\eta_{xEV}$ is the constant charge efficiency, $I_{xEV}$ is the current, and $V_{xEV}$ is the pack voltage:
\begin{equation}
    V_{xEV}(t)=V_{xEV,cell}(t) \cdot N_{xEV,s}
\end{equation}
where $N_{xEV,s}$ is the number of cells in the pack in series. The cell voltage $V_{xEV,cell}$  determined using a 0th order equivalent circuit model \cite{lam2011practical}:
\begin{equation} \label{eq:xEVVoltage}
    V_{xEV,cell}(t)=V_{OC}(SOC_{xEV})-R_0(SOC_{xEV}) \cdot I_{xEV,cell}(t)
\end{equation}
Where $I_{xEV,cell}$ is the current applied to individual cells, $R_0$ is the internal resistance of each cell and $V_{OC}$ is the open-circuit voltage. Both open circuit voltage and internal resistance are SOC dependent  \cite{lam2011practical}. The cell current is defined as
\begin{equation} \label{eq:xEVC}
    I_{xEV,cell}(t)=\frac{I_{xEV}(t)}{ N_{xEV,p}}
\end{equation}
Where $N_{ES,p}$ is the number of cells in the pack in parallel.  Finally, the SOC is
\begin{equation} \label{eq:xEVSOC}
    \frac{dSOC_{xEV}(t)}{dt}=-\frac{I_{xEV,cell}(t)}{Q_{cell}(t)}
\end{equation}
where $Q_{cell}$ is the nominal cell capacity. 

The number of cells in series and parallel are selected to meet the total pack capacity $Q_{xEV,des}$ and voltage $V_{xEV,des}$ of a specific vehicle , which is an input to the residential power demand model \cite{goutham2020machine}. The initial SOC of the vehicle at plug-in is determined as a function of vehicle efficiency and miles traveled, which are randomized together with the time of plug-in. %Additionally, the expected time of plug-in at home is captured using the EV Project data \cite{goutham2020machine}. 
The vehicle is charged using a standard level 2 charger with constant 7.56 kW and following a constant-current constant-voltage protocol. 

\subsection{User Activities} \label{subsection:UserActivities}
A number of user activities associated with non-negligible power demand are considered in this model. The user activity subsystem uses a statistical approach and reproduces the ATUS census data \cite{muratori2014dynamic,ATUS}. User activities are divided among deferrable and non-deferrable loads. The non-controllable activities include sleeping, no-power activity, cleaning, cooking, and leisure. The power request associated with each non-controllable activity is summarized in Table \ref{tab:noncontrollableAppliance}.

\begin{table}[!htb]
    \caption{Operating Power of Non-Controllable Activities}
    \label{tab:noncontrollableAppliance}
    \centering
    \resizebox{0.35\columnwidth}{!}{
    \begin{tabular}{c c}
         \hline
         \textbf{Activity} & \textbf{Power $P_{Act,NC}$ [W]}\\
         \hline
         Sleeping & 0\\
         No-Power Activity & 0\\
         Cleaning & 1250\\
         Cooking & 1225\\
         Leisure & 300\\
         \hline
    \end{tabular}}
\end{table}
Similarly, activities related to the utilization of smart appliances have been incorporated. For laundry, an all-in-one machine is considered in which washer and dryer are combined in a single unit. The power demands associated with deferrable loads are summarized in Table \ref{tab:controllableAppliance}.

\begin{table}[!htb]
    \caption{Operating Power and Duration of Controllable Appliances}
    \label{tab:controllableAppliance}
    \centering
    \resizebox{0.35\columnwidth}{!}{
    \begin{tabular}{c c c}
         \hline
         \textbf{Appliance} & \textbf{Power  [W]} & \textbf{Duration [min]}\\
         \hline
         Washer & 425 & 30 \\
         Dryer & 3400 & 30 \\
         Dishwasher & 1800 & 60 \\
         \hline
    \end{tabular}}
\end{table}

\section{Baseline Control} \label{sec:BaselineController}
To simulate the household model, a baseline controller is implemented to reproduce the scenario of no load coordination. The strategy includes a dead-band control for the HVAC and an energy storage controller that charges the battery when there is solar surplus and discharges when the power demand is not met directly from local generation. The other demands are fulfilled as soon as requested, with no deferral. %The integration of the baseline controller with the smart home plant model is illustrated in Figure \ref{fig:BaselineControl} where the case definition variables represents the external inputs depending on plant model settings, for example ambient temperature depending house location.

\subsection{HVAC Dead-Band Control}
The baseline thermostat control of HVAC aims to keep the household temperature $T_a$ within a desired bound. 
%A dead-band controller accomplishes this by only imposing on and off commands depending on ambient and home temperatures. 
%For example, in summer 
When the ambient temperature is above a certain temperature threshold, the HVAC is set to cooling operation and will turn on following:
\begin{equation} \label{eq:HVACBaselineControl}
    \pi_{cool}=
        \begin{cases}
            0 & \text{if $T_a$ $\leq$ $T_{des,c}$ - $\Delta T_{band}$}\\
            1 & \text{if $T_a$ $>$ $T_{des,c}$ + $\Delta T_{band}$}\\
        \end{cases}
\end{equation}
where $2\Delta T_{band}$ is the tolerance band, and $T_{des,c}$ is the desired temperature for cooling.Conversely,  when the ambient temperature is below a threshold value, the HVAC is on the heating mode and activated using: 
%when the household temperature goes below a set point $T_{des,h}$ with a :
\begin{equation}
    \pi_{heat}=
        \begin{cases}
            1 & \text{if $T_a$ $<$ $T_{des,h}$ - $\Delta T_{band}$}\\
            0 & \text{if $T_a$ $\geq$ $T_{des,h}$ + $\Delta T_{band}$}\\
        \end{cases}
\end{equation}
where $T_{des,h}$ is the heating setpoint.

\subsection{Energy Storage Charge and Discharge Control} 
The stationary energy storage is controlled based on the power balance of Eq. \ref{eq:PowerBalanceModel}) and the availability of surplus solar energy. The baseline controller utilizes the solar power to meet the instantaneous household power demand. If the generation is greater than the demand and the SOC is less than 90\%, the surplus power is used to charge the battery. When the solar generation is less than the total demand and the SOC of the battery is higher than 20\%, power is drawn from the energy storage system. Any additional power deficit is then covered by drawing from the electrical grid:
\begin{equation} \label{eq:ESBaselineControl}
    \pi_{ES}=
        \begin{cases}
            1 & \text{if $SOC_{ES}$ $<$ $0.9$ \& $P_{Solar}$ $\leq$ $P_{Household}$}\\
            -1 & \text{if $SOC_{ES}$ $>$ 0.2 \& $P_{Solar}$ $>$ $P_{Household}$}\\
            0 & \text{else}\\
        \end{cases}
\end{equation}
where $P_{Household}$ is the right hand side of Eq. (\ref{eq:PowerBalanceModel}). 

%\subsection{Deferrable User Activities and Vehicle Charging} 
%In the baseline scenario, deferrable user activities and vehicle charging demands are met as soon as requested. For example, if a user turns on the dishwasher, the appliance runs immediately. Similarly, if a user plugs in a vehicle at 7pm and requires 10 kWh to fully recharge, the vehicle will start charging at 7:00 pm and be completely charged by approximately 8:20 pm. 

\section{Real Time Control} \label{sec:realTimeControl}
%A multi-objective optimization problem is formulated as a decentralized model predictive control problem, which is solved using a meta-heuristic optimization algorithm. This novel HEM strategy allows for the combined optimization of the load scheduling, while minimizing the disruption to the user. Moreover, the MPC formulation compensates for the unknown future demand inputs from users. Finally, the combination of a decentralized approach based on a sequential scheme and the adoption of  a meta-heuristic optimization algorithm for solving the non-convex Mixed Integer Nonlinear Programming problem, allows for near-optimal performance and faster than real time computation. Ultimately, the objective of this study is to perform a techno-economical analysis on the deployment of HEM strategies under various conditions, such as location, household sizes and electric vehicle characteristics. For this reason, the ability of the proposed controller to achieve near-optimal solution for a variety of simulation scenario and its modularity, which allows the evaluation of different equipment and technologies, are crucial and are the main novelty of the proposed approach.

\subsection{Optimization Problem Formulation} \label{subsec:centralOptimalControl}
\begin{figure}[!htb]
 	\begin{center}
 	\includegraphics[trim=0cm 0cm 0cm 0cm, clip=true, scale=0.45]{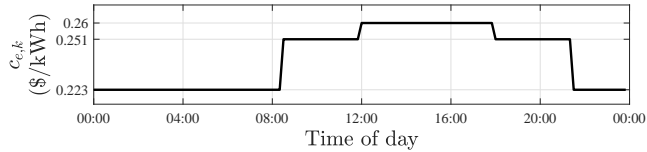}
 		\caption{TOU pricing scheme}
 		\label{fig:TOUPricing}
 	\end{center}
\end{figure}

The objective of the smart home energy management algorithm is to coordinate the scheduling of smart appliances, vehicle charging and operate the HVAC to minimize the electricity cost in response to a TOU pricing scheme, and to minimize the associated discomforts to the user.
Resident discomforts are formulated as a deferral cost associated with a delay in meeting a load demand after it was requested, and as a temperature discomfort cost associated with the temperature deviation from a desired set point. For the smart home considered in this paper, the global optimal control problem is formulated in discrete form as:
\begin{subequations} \label{eq:MINLP}
\allowdisplaybreaks
 \begin{align}
 \min_{u}J(x,u,w)= \sum_{k=0}^{N-1} &c_{e,k}P_{Grid,k}+ c_{d,k}\left[E(w_k)-u_k\right] + c_{t,k}\left[T_{a,k}-T_{set}\right]^2 \label{eq:MINLP_Objective}\\
 \text{Subject}~\text{to} \quad \quad \quad     x_{k+1}&=f_k(x_k,u_k,w_k)       \label{eq:MINLP_state}\\
    x_{min}& \leq x_k \leq x_{max}  \label{eq:MINLP_stateCon}\\
    u_{HVAC,k} &\in \{0,0.5,1\}     \label{eq:MINLP_HVAC}\\
    0 &\leq u_{xEV,k}\leq u_{max}   \label{eq:MINLP_xEV}\\
     u_{min} &\leq u_{ES,k} \leq u_{max}  \label{eq:MINLP_ES}\\
    u_{D/L,k}&\in \{0,1\}           \label{eq:MINLP_App1}\\
    u_{D/L/xEV,k} &=0 ~\forall k<E(w_k) \label{eq:MINLP_Enable}\\
    u_{D/L/xEV,k}&=0 ~\forall k>D(w_k)  \label{eq:MINLP_Deadline}\\
    \sum_{k=0}^{N-1} u_{D/L/xEV,k} &= C    \label{eq:MINLP_Complete}\\
    u_{L/D,k} &\leq 1-s_{L/D,k}      \label{eq:MINLP_slack1}\\
    u_{L/D,k-1}-u_{L/D,k} &\leq 1-s_{L/D,k} \label{eq:MINLP_slack2}\\
    s_{L/D,k} &\leq s_{L/D,k-1}      \label{eq:MINLP_slack3}\\
    \label{eq:MINLP_PowerBalance}
        P_{D,k}+P_{L,k}+P_{xEV,k}&+P_{HVAC,k}+P_{ND,k} =P_{Grid,k}+P_{ES,k}+P_{Solar,k}\\
    P_{Grid,k}& \leq P_{cap,k}      \label{eq:MINLP_Grid}
\end{align}
\end{subequations}
where $J(x,u,w)$ is the objective function, $k$ is the time step, $c_{e,k}$ is the electricity price, $c_{d,k}$ is the deferral cost that penalizes the delay in activation of an appliance after it was requested by the user, $T_{set}$ is the temperature set point and $c_{t,k}$ is the cost associated with a temperature deviation. The values of $c_{d,k}$ and $c_{t,k}$ are calibrated by the user based on their tolerance of the associated discomfort. 
In this paper, the electricity price is based on a TOU pricing scheme with $c_{e,k}$ as defined in Figure \ref{fig:TOUPricing}. The deferral cost $c_{d,k}$ is constant at $6e^{-7}$ while $c_{t,k}$ is a time varying parameter that depends on real time ambient and home temperatures.

%, as defined by Equation (\ref{eq:objective_weights}), where $T_{\infty,k}$ is the ambient temperature and $T_{ref}$ is the reference temperature to determine if it is winter or summer season.}
% \begin{comment}
% \begin{equation} \label{eq:objective_weights}
%     c_{t,k}=
%         \begin{cases}
%             1.5e^{-5} & \text{if $\overline{T}_{\infty,k}>T_{ref}$, $T_{a,k}-T_{set}>$1$\degree$C}\\
%             3e^{-10} & \text{if $\overline{T}_{\infty,k}>T_{ref}$, 0$\degree$C$\leq$T_{a,k}-T_{set}$\leq$1$\degree$C}\\
%             2e^{-10} & \text{if $\overline{T}_{\infty,k}>T_{ref}$, -1$\degree$C$\leq$T_{a,k}-T_{set}$\leq$0$\degree$C}\\
%             1e^{-5} & \text{if $\overline{T}_{\infty,k}>T_{ref}$, $T_{a,k}-T_{set}<$-1$\degree$C}\\
%             1e^{-5} & \text{if $\overline{T}_{\infty,k}$\leq$T_{ref}$, $T_{a,k}-T_{set}>$1$\degree$C}\\
%             2e^{-10} & \text{if $\overline{T}_{\infty,k}$\leq$T_{ref}$, 0$\degree$C$\leq$T_{a,k}-T_{set}$\leq$1$\degree$C}\\
%             3e^{-10} & \text{if $\overline{T}_{\infty,k}$\leq$T_{ref}$, -1$\degree$C$\leq$T_{a,k}-T_{set}$\leq$0$\degree$C}\\
%             1.5e^{-5} & \text{if $\overline{T}_{\infty,k}$\leq$T_{ref}$, $T_{a,k}-T_{set}<$-1$\degree$C}\\
%         \end{cases}
% \end{equation}
% \end{comment}
The state vector is denoted as $x$, $u$ is the control vector, and $w$ is the external inputs vector, which are defined as:
\begin{subequations} \label{eq:MINLPx}
 \begin{align}
    x_k&=[T_{a,k},SOC_{xEV,k},SOC_{ES,k}]^T \\
    u_{k}&=\left[u_{HVAC,k},u_{xEV,k},u_{D,k},u_{L,k}, u_{ES,k}\right]^T \\
    w_{k}&=\left[T_{\infty,k},P_{ND ,k},SOC_{xEV,0},P_{Solar,k}\right]^T 
    \end{align}
\end{subequations}
where $u_{HVAC}$, $u_{xEV}$, $u_D$, $u_L$ and $u_{ES}$ are the control commands for the HVAC, xEV charging, laundry, dishwasher and energy storage respectively. The system disturbances $w_k$ are the ambient temperature, the power demand of non-deferrable activities, the initial vehicle SOC $SOC_{xEV,0}$ and the solar power generation.
The system is subject to the discrete state dynamic defined in Eq. (\ref{eq:MINLP_state}), while the states are constrained based on Eq. (\ref{eq:MINLP_stateCon}). Depending on the type of appliance, the control inputs have different constraints. Specifically, the HVAC has three discrete modes of operation as defined in Eq. (\ref{eq:MINLP_HVAC}), the energy storage control input is limited by a minimum and maximum power as defined in Eq. (\ref{eq:MINLP_ES}), and the laundry and dishwasher are controlled by binary signals as defined in Eq. (\ref{eq:MINLP_App1}). All the deferrable activities can only be completed 
after a request from the user, Eq. (\ref{eq:MINLP_Enable}); must be completed before the deadline\textcolor{red}{,} Eq. (\ref{eq:MINLP_Deadline}); and the duration is limited by the completion time, Eq. (\ref{eq:MINLP_Complete}). Here, $E(w_k)$ denotes the enabling time of each appliance, $D(w_k)$ is the last possible completion time of that appliance, and $C$ is the duration of each operation.
For the dishwasher and laundry only, which are activities that cannot be interrupted and resumed once initiated, a slack variable $s$ is introduced together with the constraints defined by Equations (\ref{eq:MINLP_slack1}) - (\ref{eq:MINLP_slack3}).
The household power must satisfy the power balance defined in Eq. (\ref{eq:MINLP_PowerBalance}) and the power drawn from the grid should also be smaller than the capping power, as defined in Eq. (\ref{eq:MINLP_Grid}).
The completion time $C$ and deadline $D(w_k)$ for the deferrable appliances are summarized in Table \ref{Tab:CompletionTimeDeadlineAppliance}. There needs to be nine and six time steps for laundry and dishwasher to complete operation. The completion time for charging xEV back to the desired SOC level depends on the initial xEV SOC. All the three appliances need to complete the operation within 48 time steps after the enabling time.

\begin{table}[!htb]
\caption{Completion Time and Deadline (Time Steps) for Deferrable Appliances.}
\label{Tab:CompletionTimeDeadlineAppliance}
\centering
\resizebox{0.55\columnwidth}{!}{
\begin{tabular}{r | c | c | c} 
\toprule
\textbf{Appliances} & \textbf{Laundry} & \textbf{Dishwasher} & \textbf{xEV} \\ 
\hline \hline
Completion Time & 9 & 6 & Depends on initial SOC \\ 
\hline
Deadline & 48 & 48 & 48 \\ 
\bottomrule
\end{tabular}
}
\end{table}

Finally, $N$ represents the number of time steps over which this problem is evaluated and $\Delta t$ is the associated discretization step. The total simulation time $T_{end}$ is calculated as $T_{end}=N\cdot \Delta t$. Depending on the objective of the study and types of models used, $T_{end}$ ranges from days to months while $\Delta t$ varies from less than one second, to several seconds \cite{ghiaus2010calculation,karlsson2011application}. The values of the state and input constraints are summarized in following equations:
 \begin{align}\label{eq:stateinputlimits}
    T_{ref}&=20\degree\text{C}\\
    P_{cap}&=14\text{kW}\\
    T_{set}&=
        \begin{cases}
            22\degree\text{C},~\text{if $T_{\infty}$ $\leq$ $T_{ref}$}\\
            18\degree\text{C},~\text{if $T_{\infty}$ $>$ $T_{ref}$}\\
        \end{cases}\\
    T_{set}-1\degree\text{C} &\leq T_a \leq T_{set}+1\degree\text{C} \\
    20\% &\leq SOC_{xEV} \leq 80\%\\
    20\% &\leq SOC_{ES} \leq 80\% \\
    0\text{A} &\leq u_{xEV} \leq 2.5N_{xEV,s}\\
    0\text{A} &\leq u_{ES} \leq 2.5N_{ES,s}
\end{align}    
 The solution of the optimization problem defined in  Eq. (\ref{eq:MINLP}) requires perfect knowledge of the external input vector $w_k$ for the entire duration of the simulation, which is impossible for online implementations. To compensate for future uncertainties and  enable online implementation, the optimization problem defined in Eq. (\ref{eq:MINLP}) is solved using a receding horizon approach, where the external inputs are kept constant within the horizon  and the combination of control inputs that minimizes a performance metric while satisfying the constraint is selected \cite{richalet1978model, rawlings2017model}. In this paper, the MPC is solved using a time horizon of 8 hours, which is selected to capture the deadline of each operation. A discretization of $\Delta t = 600s$ is selected to properly capture the dynamic of the temperature in the house. 

The other advantage of a receding horizon approach is that it scales well when the total simulation time $T_{end}$ is large. 
However, the nonlinear system dynamics, both integer and continuous control variables, and the large number of states and control inputs make the optimization problem defined in Eq. (\ref{eq:MINLP}) computationally expensive for real-time implementation even for a finite time horizon $T_H$ \cite{huang2019hybrid}.
In such cases, a decentralized approach permits real-time deployment by breaking down the original large problem formulation down into multiple smaller decoupled sub-problems that can be solved sequentially \cite{christofides2013distributed}.
For this reason, a decentralized model predictive controller is proposed in this paper to achieve faster than real-time computation without compromising the optimality for the solution.  

\begin{figure*}[b]
\centering
%trim={<left> <lower> <right> <upper>}
    \includegraphics[trim={0cm 4cm 0cm 3cm},clip,scale=0.4]{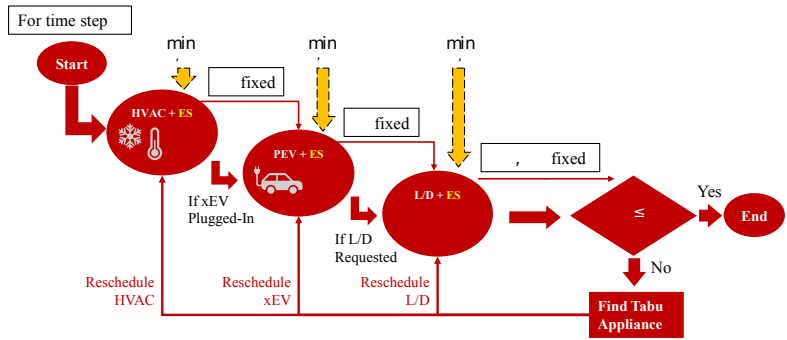}
    \caption{Hierarchical optimization scheme for the sub-problems of the HEM strategy at one time step}
    \label{fig:HierarchicalMPC}
\end{figure*}

\subsection{Sequential Scheme for Decentralized Load Coordination}\label{sequential subproblems}
The centralized optimization problem defined in Eq. (\ref{eq:MINLP}) is divided into three equivalent sub-problems based on the frequency of each system requests and activation. In this case the three problems are the optimization of the HVAC system and energy storage; the vehicle charging and energy storage; and laundry, dishwasher and energy storage. 
The scheduling of laundry and dishwasher are combined due to their similarities in constraints and input variable types, while the energy storage system is considered in each sub-problem because it affects the amount of energy available to the other controlled appliances, thereby affecting their scheduling, which indirectly couples the appliances. The other indirect coupling among the sub-problems is the capping constraint defined in Eq. (\ref{eq:MINLP_Grid}), but since this capping constraint is violated infrequently, its satisfaction is verified after the completion of the sequence of optimizations. The structure, hierarchy and interactions between the sub-problems is shown in Figure \ref{fig:HierarchicalMPC}. 

\subsubsection{HVAC and Energy Storage} \label{subsubsec:HVACESSubOpt}
The first sub-problem optimizes the operation of HVAC and energy storage over the time horizon $T_H$, which is divided in to $N_H$ equal time steps. The global cost function defined in Eq. (\ref{eq:MINLP_Objective}), is then modified to consider the electricity cost associated with the heating and cooling and the temperature deviation only. The other appliance loads are neglected, simplifying the state dynamics to $f_k^{(1)}$, state vector $x_k^{(1)}$, control vector $u_k^{(1)}$ and limiting the number of required constraints. 

The first sub-problem is:
\begin{subequations} 
\begin{equation} \label{eq:HVACsubObj}
\min_{u}J(x,u,w)=\sum_{k=0}^{N_{H}-1} c_{e,k}P_{HVAC,k} + c_{t,k}\left[T_{a,k}-T_{set}\right]^2
\end{equation}
 \text{Subject to:}
 \begin{align}
    x_{k+1}^{(1)}&=f_k^{(1)}(x_k^{(1)},u_k^{(1)},w_k^{(1)}) \\
    x_{min}^{(1)} &\leq x_k^{(1)} \leq x_{max}^{(1)}\\
    u_{HVAC,k} &\in \{0,0.5,1\}\\
    u_{ES,min} &\leq u_{ES,k} \leq u_{ES,max}\\
    \begin{split}
        P_{HVAC,k}&+P_{ND,k}\\&=P_{Grid,k}+P_{ES,k}+P_{Solar,k}
    \end{split}
\end{align}
\end{subequations}

Solving the above sub-problem yields the control commands $u_{HVAC}^{(1)}\in \mathbb{R}^{N_H}$ for HVAC and $u_{ES}^{(1)}\in \mathbb{R}^{N_H}$ for ES over the time horizon $T_H$. The optimal HVAC command $u_{HVAC}^{(1)}$ and associated power demand $P_{HVAC}^{(1)}\in \mathbb{R}^{N_H}$ over the horizon $T_H$ are then fed to the next sub-problem as external inputs.

\subsubsection{Vehicle Charging and Energy Storage} 
If xEV is plugged in by the user, the second optimization sub-problem optimizes the schedule of the xEV charging and energy storage. Conversely, if there is not vehicle plugged in, this step is skipped as the solution is trivial. 
In this optimization, HVAC command is fixed as $u_{HVAC}^{(1)}$ over $T_H$ and the corresponding power demand is known $P_{HVAC}^{(1)}$ and fixed. Similar to the prior case, the global objective function is modified to only include the cost of electricity drawn from the grid and associated with the vehicle charging and the deferral cost. The other appliance loads are neglected, simplifying the state dynamics to $f_k^{(2)}$, state vector $x_k^{(2)}$, control vector $u_k^{(2)}$ and limiting the number of required constraints. The optimization problem becomes: 
\begin{subequations} 
 \begin{align}
\min_{u}J(x,u,w)&=\sum_{k=0}^{N_{H}-1} c_{e}(k)P_{Grid,k}+c_{d,k}\left[E(w_k)-u_{xEV,k}\right]\\
\text{Subject~to}: x_{k+1}^{(2)}&=f_k^{(2)}(x_k^{(2)},u_k^{(2)},w_k^{(2)}) \\
    x_{min}^{(2)} &\leq x_k^{(2)} \leq x_{max}^{(2)} \label{eq:MILP_S2_C1}\\
    u_{min}^{(2)} &\leq u_{xEV,ES,k} \leq u_{max}^{(2)} \\
    \begin{split}     \label{eq:MINLP_Constraint_4_sub2}
        P_{xEV,k}&+P_{HVAC,k}^{(1)}+P_{ND,k}\\
        &=P_{Grid,k}+P_{ES}(k)+P_{Solar,k}  \end{split}\\
    u_{xEV,k}&=0 ~\forall k<E(w_k) \label{eq:MILP_S2_C2}\\
    u_{xEV,k}&=0, ~\forall k>D(w_k)\\
    \sum_{k=0}^{N_H-1} u_{xEV}(k)& = C\label{eq:MILP_S2_C3}
\end{align}
\end{subequations}
By including the constraint in Eq. (\ref{eq:MINLP_Constraint_4_sub2}), the controller will account for both energy generation and stored when deciding to draw power from the grid. This is the reason for using $P_{Grid,k}$ in the cost function in place of $ P_{xEV,k}$. The results of the this optimization problem is the optimal control commands for xEV $u_{xEV}^{(2)}\in \mathbb{R}^{N_H}$ and energy storage $u_{ES}^{(2)}\in \mathbb{R}^{N_H}$.

\subsubsection{Laundry, Dishwasher, and Energy Storage} 
If laundry or dishwasher is requested by the user, the third optimization sub-problem is solved. If the user has not requested any appliance, this problem is skipped as the result is trivial. As for the two previous problems, the cost function, state dynamic, control inputs, external inputs and constraints are all modified to represent the scheduling of the two non-interruptible appliances. In this third level of the optimization, the control commands for the HVAC and vehicle charging  over $T_H$ are $u_{HVAC}^{(1)}$ and xEV $u_{xEV}^{(2)}$.
The third sub-problem is:
\begin{subequations} \label{eq:MINLP_sub3}
 \allowdisplaybreaks
  \begin{align}
 \label{eq:MINLP_Objective_sub3}
\min_{u}J(x,u,w)=\sum_{k=0}^{N_H-1} &c_{e,k}P_{Grid,k}+ c_{d,k}\left[E(w_k)-u_{D/L,k}\right]\\
 \text{Subject}~\text{to}:    x_{k+1}^{(3)}&=f_k^{(3)}(x_k^{(3)},u_k^{(3)},w_k^{(3)})  \\
    x_{min}^{(3)}& \leq x_k^{(3)} \leq x_{max}^{(3)} \\
    u_{D/L,k}&\in \{0,1\}          \\
    u_{D/L,k} &=0 ~\forall k<E(w_k) \\
    u_{D/L,k}&=0 ~\forall k>D(w_k) \\
    \sum_{k=0}^{N_H-1} u_{D/L,k} &= C   \\
    u_{L/D,k} &\leq 1-s_{L/D,k}   \\
    u_{L/D,k-1}-u_{L/D,k} &\leq 1-s_{L/D,k}\\
    s_{L/D,k} &\leq s_{L/D,k-1}   \\
    \begin{split} 
        P_{D,k}+P_{L,k}+P_{xEV,k}^{(2)}&+P_{HVAC,k}^{(1)}+P_{ND,k} \\=P_{Grid,k}+P_{ES,k}&+P_{Solar,k}
    \end{split}  
\end{align}
\end{subequations}
The solution provides the optimal control commands for laundry $u_L^{(3)}\in \mathbb{R}^{N_H}$, dishwasher $u_D^{(3)}\in \mathbb{R}^{N_H}$, and energy storage $u_{ES}^{(3)}\in \mathbb{R}^{N_H}$. Again, the total grid power is used instead of the power associated with the appliances so that the effect of the energy storage and local generation is caputred.

\subsubsection{Power Capping Check} 
After determining the optimal control commands for HVAC $u_{HVAC}^{(1)}$, vehicle charging $u_{xEV}^{(2)}$, laundry $u_L^{(3)}$, dishwasher $u_D^{(3)}$ and energy storage $u_{ES}^{(3)}$ using the sequential scheme, the total power request from the grid is calculated using Eq. (\ref{eq:MINLP_PowerBalance}) and its value is compared against the maximum allowable grid power defined in Eq. (\ref{eq:MINLP_Grid}). If the constraint is violated at any time $k$ in $N_H-1$, the algorithm finds the appliance with largest power demand during the constraint violation window. This appliance is referred to as the Tabu Appliance in Figure \ref{fig:HierarchicalMPC}. The sequential optimization is performed again, but this time an additional constraint is added to the sub-problem associated with the Tabu Appliance that prevents the specific appliance to be utilized during those time steps. The sequential optimization is repeated until the constraint is satisfied.

The finalized optimal control commands are denoted as 
\begin{equation}   
\begin{bmatrix}
u_{HVAC}^{*},~u_{xEV}^{*},~u_{L}^{*},~ u_{D}^{*},~u_{ES}^{*}
\end{bmatrix}
\in \mathbb{R}^{5\times N_H}
\end{equation}.

\subsubsection{Shrinking Horizon Implementation}

A shrinking time horizon ensures that every demanded activity is completed within the simulation time \cite{ma2012demand}.
Thus, as the simulation progresses towards the final time step $T_{end}$, specifically when $T(k)+T_H>T_{end}$, the time horizon is shrunk by replacing $T_H$ with $T_{H,Sh}$:
\begin{equation}\label{eq:recedinghorizon}
    T_{H,Sh} = T_{end}-T(k), \text{if}~T(k)+T_H>T_{end}
\end{equation}

\subsection{Solution via Genetic Algorithm} 

Solving the sub-problems of the sequential controller to obtain the optimal solution over the horizon $T_H$ requires significant computational effort. %due to the presence of \textcolor{red}{multiple appliances with different types of command inputs and constraints.}
% \st{Each sub-problem requires to determine an optimal control sequence over the receding horizon $T_H$.} 
To ensure that the computation is performed faster than real time, a meta-heuristic algorithm is used. Among meta-heuristic approaches, GA provides near optimal solutions in problems with integer commands \cite{liu2020multi}. For each sub-problem described in section \ref{sequential subproblems}, a population of feasible control commands is first initialized, where each individual in the population represents a candidate solution for the appliance command over $T_H$ \cite{passino2005biomimicry}.
The objective value of all individuals is then evaluated and the individuals are ranked in order of lowest cost. The next population of candidate solutions is initialized with a fixed number of lowest cost individuals. These parent individuals are then modified, using both mutation and crossover, to generate the children and compose the new generation. The fraction of new children constructed by mutation and crossover are predefined parameters in the GA algorithm, as summarized in Table \ref{table.GeneticAlgorithmSetting}. 

\begin{table}[!htb]
\caption{GA Settings for Sub-Problem Optimization}
\label{table.GeneticAlgorithmSetting}
\centering
\resizebox{0.55\columnwidth}{!}{
\begin{tabular}{l | c | c | c | c}
			\toprule[0.75pt]
			\multirow{2}{*}{GA Options} & \multicolumn{2}{c}{HVAC} & \multirow{2}{*}{xEV} & \multirow{2}{*}{L/D}  \\
			\cmidrule(lr){2-3}  & Summer & Winter & & \\ 
			\hline \hline
			Maximum generation & 500 & 500 & 500 & 500 \\
    \hline
			Maximum stalled generation & 50 & 50 & 50 & 50 \\
    \hline
			Elite count & 10 & 10 & 15 & 20 \\
    \hline
			Objective tolerance & 1e-2 & 1e-3 & 1e-4 & 1e-3 \\
    \hline
			Time limit [s] & 30 & 30 & 30 & 30 \\
    \hline
			Population size & 100 & 50 & 250 & 250 \\
    \hline
			Crossover fraction & 0.4 & 0.2 & 0.4 & 0.2 \\
			\bottomrule[0.75pt]
		\end{tabular}
}
\end{table}

\begin{algorithm}
\caption{Genetic Algorithm for HVAC Sub-Problem}\label{algorithm_GA}
\begin{algorithmic}[1]
\STATE \textbf{Inputs}
\STATE Obtain parameters: electricity price $c_{e,k}$, multi-objective function weights $c_{d,k}$ and $c_{t,k}$.
\STATE Non-controllable user activities, external inputs including environment and household characteristics, limits and constraints on states and inputs.
\STATE Select GA parameters based on appliance being optimized, using Table \ref{table.GeneticAlgorithmSetting}.
\STATE \textbf{Initialization}
\STATE Randomly generate the initial population of appliance commands $[u_{HVAC,k}^T, u_{ES,k}^T]$ based on selected \textit{creation function}, $k$ is the time step and $1\leq k \leq T_{H,Sh}$.
\STATE \textit{generation count} := 1
\STATE \textbf{Iteration}
\WHILE{``termination criteria of GA is not met'' }
\STATE \textbf{Evaluation}
\STATE Evaluate the ﬁtness value for each individual of the population. This involves the trade-off between grid electricity cost, delay cost, and temperature discomfort cost, $J(x,u_{HVAC,k}^T,u_{ES,k}^T,w)$ using Eq. (\ref{eq:HVACsubObj}).
\STATE \textbf{Evolution}
\STATE \textit{selection function}: assign parents as elite individuals of the current generation.
\STATE \textit{crossover function}: generate next generation children by crossover between parents.
\STATE \textit{mutation function}: generate next generation children by mutation of parent.
\ENDWHILE
\STATE Select command $u_{HVAC}^{(1)}, u_{ES}^{(1)}$ with the best fitness value in the final population.
\end{algorithmic}
\end{algorithm}
Population evolution is repeated until a defined stopping criteria is met based on the time limit, number of generations, and number of ``stalled'' generations where there was no improvement in objective value.
For the three sub-problems and depending on seasonality, the parameters are as summarized in Table \ref{table.GeneticAlgorithmSetting} and the GA is summarized in Algorithm \ref{algorithm_GA}. 
%The algorithm is implemented using the Global Optimization Toolbox of Matlab, making use of the ``\textit{gacreationuniformint}'' function for population initialization, ``\textit{selectionremainder}'' for elite selection, ``\textit{crossoverlaplace}'' for crossover operations and ``\textit{mutationpower}'' for mutation.

In the HVAC and ES optimization sub-problem, each GA individual represents their respective control commands given by $[u_{HVAC,k}^T, u_{ES,k}^T]$ with a command vector length of $2N_H$. 
When the convergence criteria is met, the individual with the lowest objective function value is saved and the vector entries of this individual are assigned as the control commands for HVAC $u_{HVAC}^{(1)}$ and energy storage $u_{ES}^{(1)}$. 

To reduce the dimension of the xEV charge scheduling sub-problem, each individual in the population represents the ES command and the time of xEV charging instead of the xEV commands for the entire time horizon. 
The times of activation vector $v_{xEV,GA}$ stores the $C$ discrete time steps $\{k_1,k_2,...,k_C\}$ in the time horizon when the xEV charging command is ``on''.
$C$ is calculated to satisfy the state constraint (\ref{eq:MILP_S2_C1}),  SOC limit constraint (\ref{eq:MILP_S2_C1}), enabling constraint (\ref{eq:MILP_S2_C2}) and completion constraint (\ref{eq:MILP_S2_C3}).
The commands are ``off'' at the other $N_H-C$ time steps. 
Through this change of variable, the total length of the individual for the vehicle and energy storage charging is reduced from $2N_H$ to $N_H+C$.
\begin{subequations} \label{eq:xEVGAInterpretation}
 \allowdisplaybreaks
\begin{align}
    \label{eq:GACommandxEV}
    v_{xEV,GA}=[k_1,k_2,\dotsc,k_C]^T
\end{align}
\begin{equation} \label{eq:OriginalCommandxEV}
u_{xEV,k}=
\begin{cases}
            1,~ \forall k\in v_{xEV,GA}\\
            0,~ \forall k\notin v_{xEV,GA}\\
        \end{cases}
\end{equation}
\end{subequations}

Finally, in the laundry and dishwasher sub-problem, a similar transformation to the xEV is used for the appliance commands:
\begin{subequations} \label{eq:LDGAInterpretation}
 \allowdisplaybreaks 
\begin{equation}
    \label{eq:GACommandL}
    v_{L,GA}=k_L
\end{equation}
\begin{equation} \label{eq:OriginalCommandL}
u_{L}(k)=
\begin{cases}
            1,~ k\in \{k_L,k_L+1,...,k_L+C_L-1\}\\
            0,~ k \notin \{k_L,k_L+1,...,k_L+C_L-1\}\\
        \end{cases}
\end{equation}
which allows the use of a horizon of length $N_H+2$.
\begin{equation}
    \label{eq:GACommandD}
    v_{D,GA}=k_D
\end{equation}
\begin{equation} \label{eq:OriginalCommandD}
u_{D}(k)=
\begin{cases}
            1,~ k\in \{k_D,k_D+1,...,k_D+C_D-1\}\\
            0,~ k \notin \{k_D,k_D+1,...,k_D+C_D-1\}\\
        \end{cases}
\end{equation}
\end{subequations}

The GA individuals $v_{L,GA}$ and $v_{D,GA}$ for laundry and dishwasher store the activation time steps of the two appliances and once initialized, they remain active for following $C_L$ and $C_D$ time steps, which are the completion time steps for laundry and dishwasher, as defined in Table \ref{Tab:CompletionTimeDeadlineAppliance}.

\section{Simulation Scenarios and Metrics}
\subsection{Case Studies}
An extensive simulation campaign is conducted to study the performance of the HEM strategy under variations of house location, seasonality, house size, electric vehicle and energy storage battery sizes. A summary of the considered locations, house sizes, xEV and ES battery sizes are summarized in Table \ref{Tab:CaseStudies}.

\begin{table}[!htb]
\caption{Case Studies Summary}
\label{Tab:CaseStudies}
\centering
\resizebox{0.7\columnwidth}{!}{
\begin{tabular}{|c|c|c|c|c|c|}
\hline
\rowcolor[HTML]{656565} 
{\color[HTML]{FFFFFF} \textbf{\begin{tabular}[c]{@{}c@{}}Case\end{tabular}}} &
  {\color[HTML]{FFFFFF} \textbf{Location}} &
  {\color[HTML]{FFFFFF} \textbf{\begin{tabular}[c]{@{}c@{}}House Size\\ (ft$^2$)\end{tabular}}} &
  {\color[HTML]{FFFFFF} \textbf{\begin{tabular}[c]{@{}c@{}}xEV Battery \\ Size (kWh)\end{tabular}}} &
  {\color[HTML]{FFFFFF} \textbf{\begin{tabular}[c]{@{}c@{}}ES Battery \\ Size (kWh)\end{tabular}}} &
  {\color[HTML]{FFFFFF} \textbf{\begin{tabular}[c]{@{}c@{}}ES Temp. \\ Control\end{tabular}}} \\ \hline
\rowcolor[HTML]{EFEFEF} 
1 &
  Columbus &
  1500-2500 &
  60 &
  14 &
  Yes \\ \hline
2 &
   &
  \textbf{500-1500} &
   &
  14 &
   \\ \cline{1-1} \cline{3-3} \cline{5-5}
3 &
   &
  \textbf{2500-3500} &
   &
  14 &
   \\ \cline{1-1} \cline{3-3} \cline{5-5}
4 &
  \multirow{-3}{*}{Columbus} &
  \textbf{3500-4500} &
  \multirow{-3}{*}{60} &
  28 &
  \multirow{-3}{*}{Yes} \\ \hline
\rowcolor[HTML]{EFEFEF} 
5 &
  \textbf{Los Angeles} &
  \cellcolor[HTML]{EFEFEF} &
  \cellcolor[HTML]{EFEFEF} &
  \cellcolor[HTML]{EFEFEF} &
  \cellcolor[HTML]{EFEFEF} \\ \cline{1-2}
\rowcolor[HTML]{EFEFEF} 
6 &
  \textbf{San Antonio} &
  \cellcolor[HTML]{EFEFEF} &
  \cellcolor[HTML]{EFEFEF} &
  \cellcolor[HTML]{EFEFEF} &
  \cellcolor[HTML]{EFEFEF} \\ \cline{1-2}
\rowcolor[HTML]{EFEFEF} 
7 &
  \textbf{Boston} &
  \multirow{-3}{*}{\cellcolor[HTML]{EFEFEF}1500-2500} &
  \multirow{-3}{*}{\cellcolor[HTML]{EFEFEF}60} &
  \multirow{-3}{*}{\cellcolor[HTML]{EFEFEF}14} &
  \multirow{-3}{*}{\cellcolor[HTML]{EFEFEF}Yes} \\ \hline
8 &
   &
   &
  \textbf{25} &
   &
   \\ \cline{1-1} \cline{4-4}
9 &
  \multirow{-2}{*}{Columbus} &
  \multirow{-2}{*}{1500-2500} &
  \textbf{100} &
  \multirow{-2}{*}{14} &
  \multirow{-2}{*}{Yes} \\ \hline
\rowcolor[HTML]{EFEFEF} 
10 &
  \cellcolor[HTML]{EFEFEF} &
  2500-3500 &
  \cellcolor[HTML]{EFEFEF} &
  \textbf{28} &
  \cellcolor[HTML]{EFEFEF} \\ \cline{1-1} \cline{3-3} \cline{5-5}
\rowcolor[HTML]{EFEFEF} 
11 &
  \multirow{-2}{*}{\cellcolor[HTML]{EFEFEF}Columbus} &
  3500-4500 &
  \multirow{-2}{*}{\cellcolor[HTML]{EFEFEF}60} &
  \textbf{14} &
  \multirow{-2}{*}{\cellcolor[HTML]{EFEFEF}Yes} \\ \hline
12 &
  Columbus &
  1500-2500 &
  60 &
  14 &
  No \\ \hline
\end{tabular}%
}
\end{table}

%%%%%%%%%%%%%%%%%%%%%%%%%%%%%%%%%%%%%%%%%%%%%%%%%%%%%%%%%%%%%%%%%%%%%%%%%%%%%%%%%%%%%%%%%%%%%%%%%%%%%%%%%%%%%%%%%%%%%%%%%%
\subsection{Metrics}\label{subsec:Metrics}
The performance evaluation of the HEM strategy against the baseline controller is performed using the following metrics.

\subsubsection{Cumulative Performance Metrics}
First, the yearly operation of the HEMS is evaluated considering the objectives in the optimization problem, namely the overall grid cost, the temperature discomfort and the appliance delay time as defined in Eq. \eqref{eq:MINLP_Objective}. These metrics are computed to understand the benefits to the user, while providing a quantification of the discomfort. Moreover, they provide an empirical verification on the optimality of the solution.

\subsubsection{Appliance Load Deferral}
\begin{figure}[!htb]
 	\begin{center}
 	\includegraphics[trim=0cm 0cm 0cm 0cm, clip=true, scale=0.6]{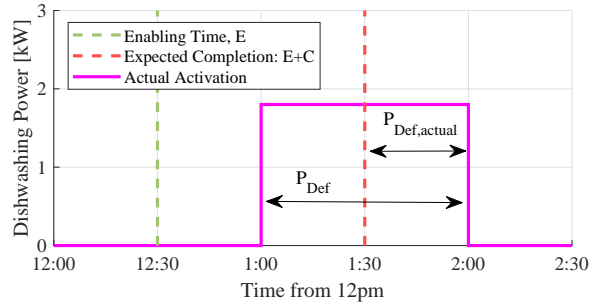}
 		\caption{Example Dishwasher Deferred Power}
 		\label{fig:DeferredPowEx}
 	\end{center}
\end{figure}

The main limitation in using the overall cost savings as a metric for the HEMS is that the improvements are subject to the specific pricing scheme considered. To understand the ability of the HEMS to schedule flexible loads independently of the relative costs, two new metrics are introduced for the first time in this paper.
First, the total deferred power $P_{Def,actual}$ is defined by calculating how much of the requested flexible power has been deferred. An example of this metric is shown in  Figure \ref{fig:DeferredPowEx} for the example of a dishwasher operation and where $P_{Def}$ is the flexible power needed to complete the operation.

 The second metric that is also  related to the scheduling performance of the HEMS is the deferral efficiency $\eta_{Def}$, defined as the percentage of the actually deferred loads $P_{Def,actual}$ in the total deferrable loads $P_{Def}$:

\begin{equation} \label{DefEff}
    \eta_{Def}= \dfrac{P_{Def,actual}}{P_{Def}} \cdot 100\%
\end{equation}

These two metrics will provide an insight on how much load the HEMS is able to shift.

\subsubsection{Solar Power Utilization and Energy Storage}
Finally, to evaluate the effect of the presence of the stationary energy storage system on the HEMS performance, the renewable energy split is analyzed. In this case, the split consists in how much of the renewable energy generation is used to instantaneously meet the household power demand, and how much is instead stored in the stationary battery. 

\section{Simulation Results}
\subsection{Effect of House Size on Controller Performance}
\begin{figure}[b]
    \centering
  \subfloat[Costs to the user \label{Fig:EffectHouseSize_CumulativePerformance}]{%
       \includegraphics[trim=0cm 0cm 0cm 0cm, clip=true, width=0.47\textwidth]{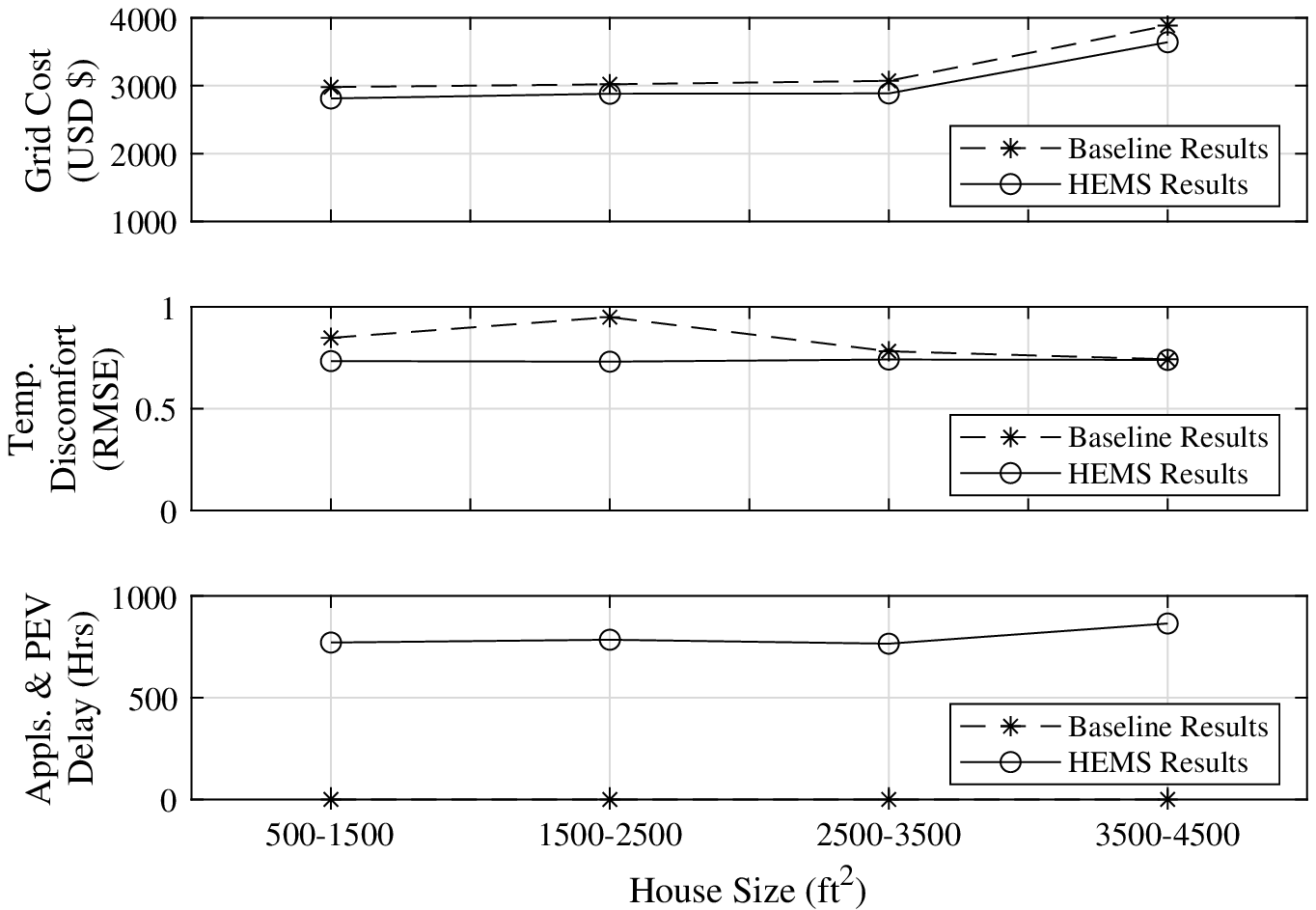}}
    \hspace{.02\linewidth}
  \subfloat[Power deferral efficiency \label{Fig:EffectHouseSize_PowerDeferral}]{%
        \includegraphics[trim=0cm 0cm 0cm 0cm, clip=true, width=0.47\textwidth]{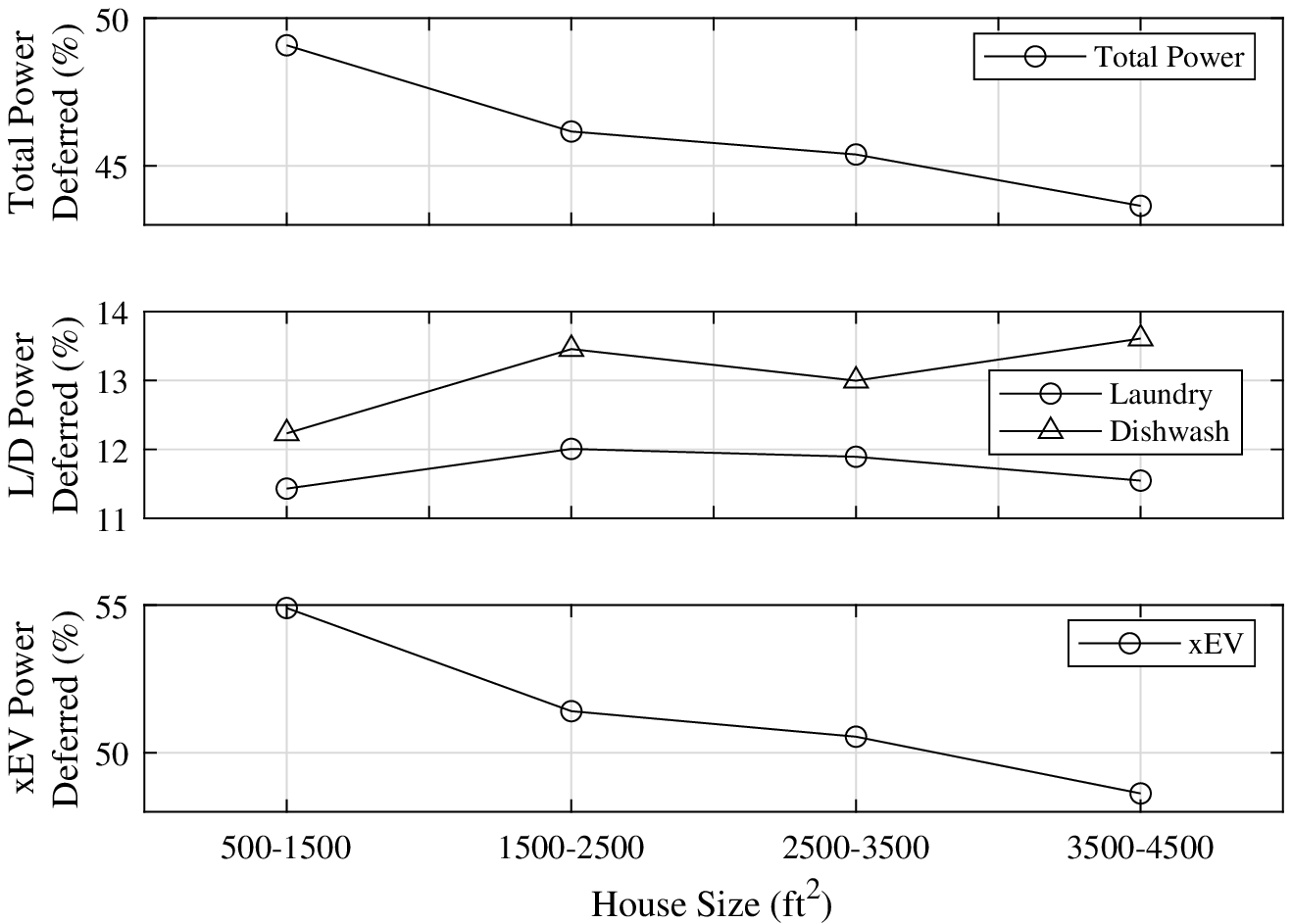}}
  \caption{Effect of household size}
\end{figure}

The effect of house size on the HEMS performance is analyzed for cases 1 to 4 of Table \ref{Tab:CaseStudies}, where the same location is considered.
The cumulative performance metrics as function of house size are shown in Figure \ref{Fig:EffectHouseSize_CumulativePerformance}. Results show that the average total grid cost savings obtained by HEMS compared to the baseline controller is approximately 6\%  and is insensitive to house size. As expected, the HEM strategy obtains improved temperature control in smaller houses, which are characterized by a faster temperature dynamics and therefore a predictive strategy is effective compared to a rule-based approach.
Moreover, there is a consistent time delay in the operation of appliance and vehicle charging.

The appliance load deferral metrics are shown in Figure \ref{Fig:EffectHouseSize_PowerDeferral}. The results indicate that smaller homes have a higher deferral efficiency compared to larger households. This is because smaller homes have a larger relative xEV load, as the total power in these homes is limited. On the other hand, the deferral efficiency tends to decrease with larger house sizes. This is because larger house sizes require more power to operate the HVAC, which reduces the load flexibility, as shown in power summary for individual appliances in Table \ref{Tab:PowSum}.
Independently from house size the HEMS deferral efficiency is always larger than 43\%, showing that the optimal strategy is delaying a large portion of the flexible power demand. A summary of the deferring performance as function of home size over one year of operation is shown in Table \ref{Tab:PowSum_Deferrable}.

\begin{table}[t]
\caption{Summary of the Annual Deferrable Power Demand Depending on Size of the Household.}
\label{Tab:PowSum_Deferrable}
\centering
\resizebox{0.7\columnwidth}{!}{
\begin{tabular}{r | c | c | c | c } 
\toprule
\textbf{House Size [ft\textsuperscript{2}]} & \textbf{500-1500} & \textbf{1500-2500} & \textbf{2500-3500} & \textbf{3500-4500}  \\ 
\hline \hline
Vehicle Charging [MWh] & 24.09 & 24.09 & 24.09 & 22.89 \\ 
\hline
Laundry [MWh] & 0.49 & 0.49 & 0.49 & 0.49 \\ 
\hline
Dishwasher [MWh] & 0.20 & 0.20 & 0.20 & 0.20 \\ 
\hline
Total Deferrable [MWh] & 24.77 & 24.77 & 24.77 & 23.58 \\ 
\hline
Deferred [MWh] & 12.16 & 11.43 & 11.24 & 10.29 \\ 
\hline 
Deferral Efficiency [\%] & 49.08 & 46.16 & 45.38 & 43.64 \\
\bottomrule
\end{tabular}
}
\end{table}

\begin{figure}[b]
    \centering
  \subfloat[Solar Power Utilization Split \label{SubFig:SolarUt}]{%
       \includegraphics[trim=0cm 0cm 0cm 0cm, clip=true, width=0.45\textwidth]{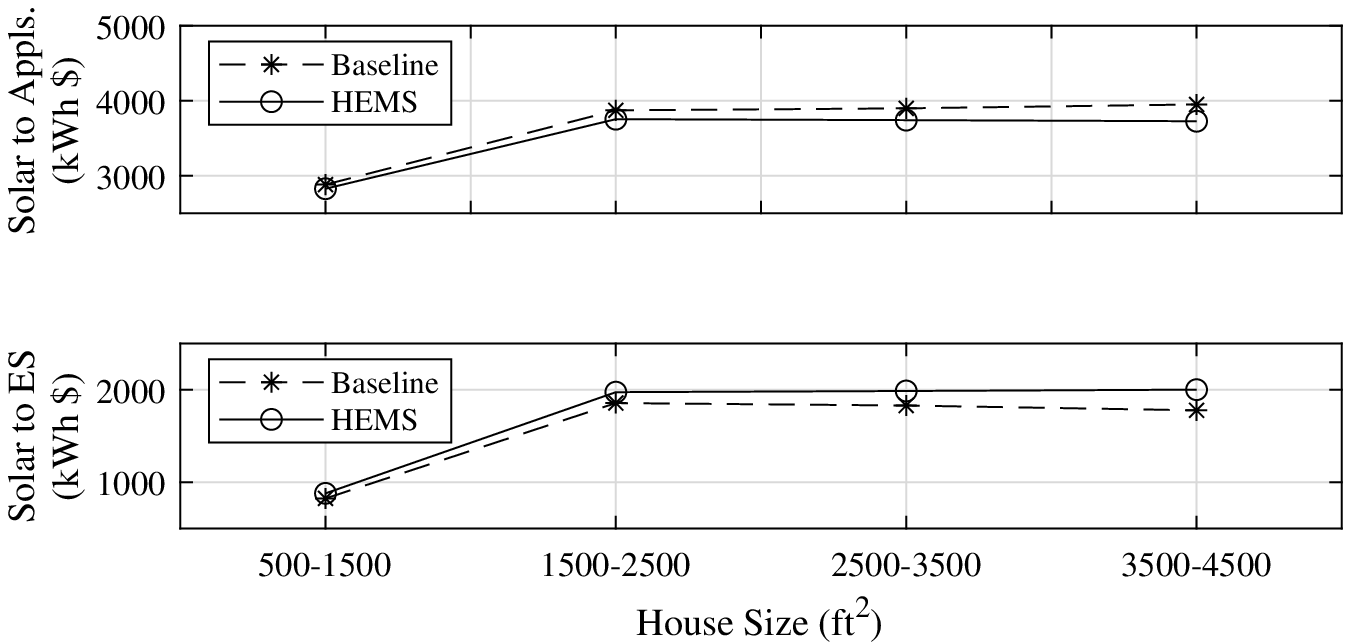}}
    \hspace{.02\linewidth}
  \subfloat[Energy Storage Behavior \label{SubFig:ESbeh}]{%
        \includegraphics[trim=0cm 0cm 0cm 0cm, clip=true, width=0.45\textwidth]{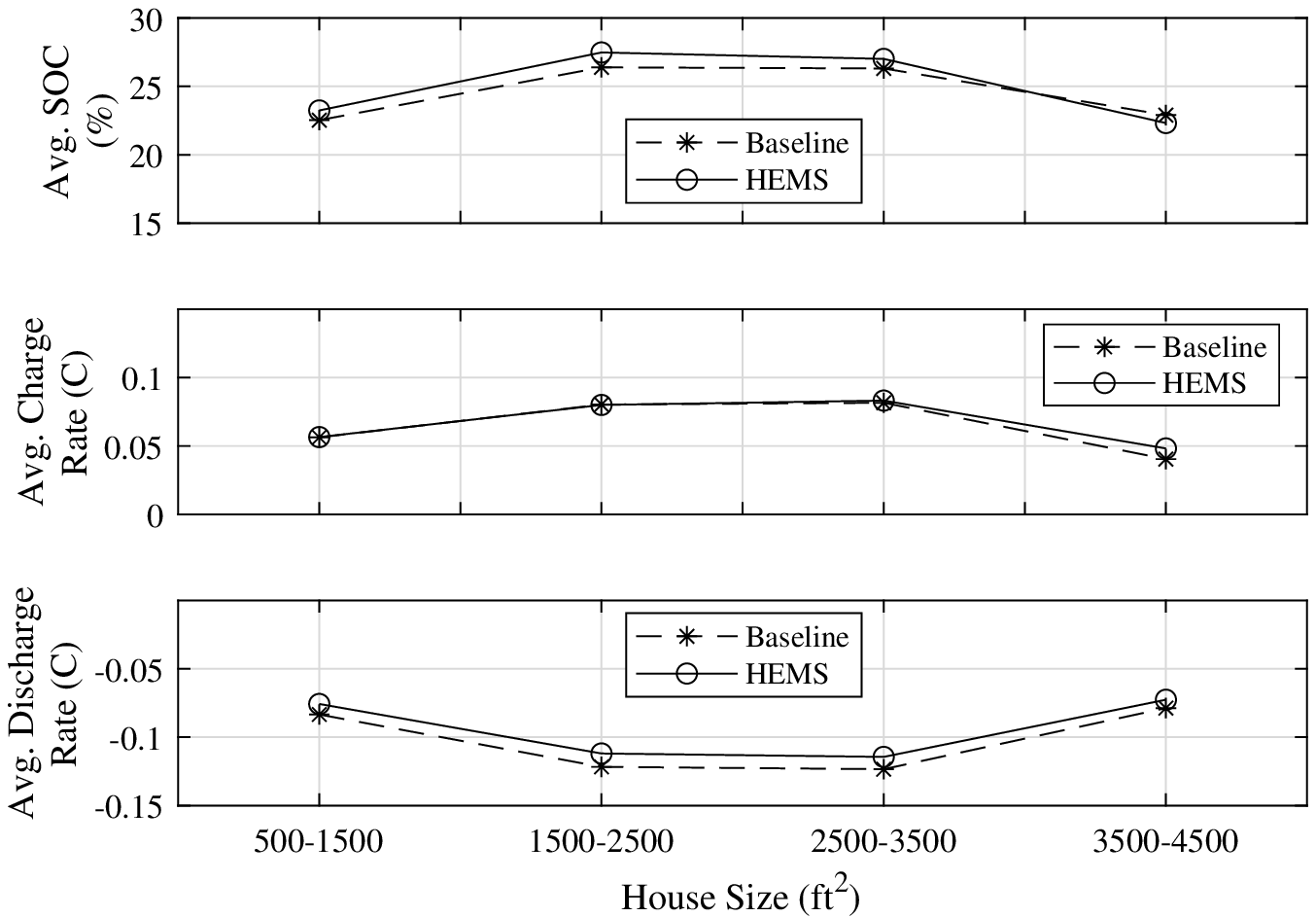}}
  \caption{Solar Power Utilization and Energy Storage Depending on Size of the Household}
  \label{Fig:EffectHouseSize_SolarEnergyStorage}
\end{figure}

\begin{table}[t]
\caption{Summary of the Annual Power Demand Depending on Size of the Household.}
\label{Tab:PowSum}
\centering
\resizebox{0.7\columnwidth}{!}{
\begin{tabular}{r | c | c | c | c } 
\toprule
\textbf{House Size [ft\textsuperscript{2}]} & \textbf{500-1500} & \textbf{1500-2500} & \textbf{2500-3500} & \textbf{3500-4500}  \\ 
\hline \hline
HVAC [MWh] & 2.86 & 4.57 & 4.86 & 8.04 \\ 
\hline
Non-Controllable Act [MWh] & 7.92 & 7.92 & 7.92 & 7.92 \\ 
\hline
Total Non-Deferrable [MWh] & 10.78 & 12.49 & 12.78 & 15.96 \\ 
\hline
Total Deferrable [MWh] & 24.77 & 24.77 & 24.77 & 23.58 \\ 
\hline
Total Power [MWh]  & 35.55 & 37.26 & 37.55 & 39.54 \\
\bottomrule
\end{tabular}
}
\end{table}
Finally, the energy storage utilization and power split metrics are shown in Figure \ref{Fig:EffectHouseSize_SolarEnergyStorage}. The baseline controller uses the majority of the solar generation to meet the power demand instantaneously. Conversely, because the HEM strategy includes a prediction of future states, constraints and costs, a larger portion of the solar power is stored in the battery compared to the baseline controller.

Compared to baseline control, the solar power stored in the stationary energy storage increases for increasing house size increases. Moreover, Figure \ref{SubFig:ESbeh} shows that HEMS results in higher average SOC in smaller homes.

\begin{figure}[t]
    \centering
  \subfloat[1606 $ft^2$ \label{SubFig:Case1_ES_SOC}]{%
       \includegraphics[trim=1cm 0cm 1cm 0cm, clip=true, width=0.48\textwidth]{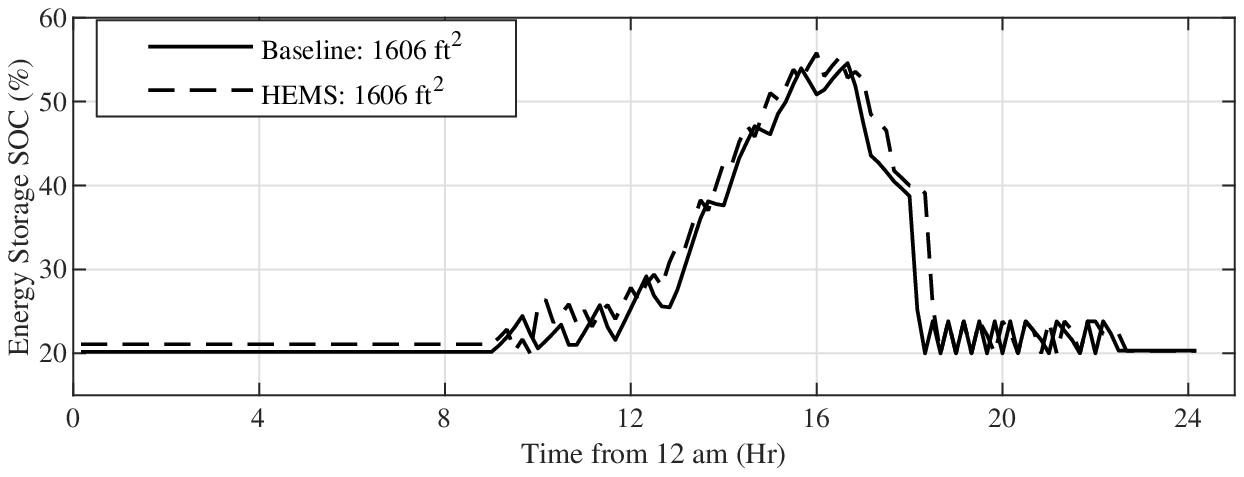}}
    \hspace{.02\linewidth}
  \subfloat[4252 $ft^2$ \label{SubFig:Case4_ES_SOC}]{%
        \includegraphics[trim=1cm 0cm 1cm 0cm, clip=true, width=0.48\textwidth]{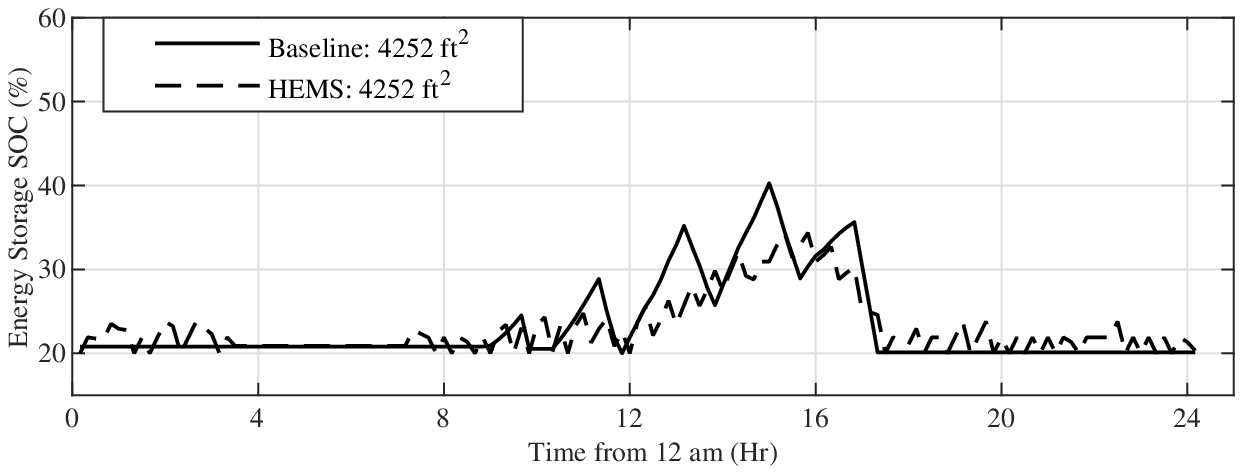}}
  \caption{Energy Storage SOC in Two Controls and Two Household Sizes}
  \label{Fig:ESSOC}
\end{figure}

Figures \ref{SubFig:Case1_ES_SOC} and \ref{SubFig:Case4_ES_SOC} illustrate the SOC of the energy storage for the mid-sized home at 1606 ft\textsuperscript{2} and the largest house size at 4252 ft\textsuperscript{2} respectively for a single day. 

For both house sizes, the frequency of charge and discharge events is higher for the HEM strategy compared to the baseline control. 
However, for the larger house, the average SOC appears to be lower for both baseline and HEM strategy controllers.
This is because the size of the stationary battery depends on the house size. Hence, the SOC changing rate is lower in the largest house. When comparing the absolute ampere-hour throughput of the two scenarios, the larger house size still achieves a higher overall utilization of the energy storage, as illustrated in Table \ref{AhThrough}. 

\begin{table}[t]
\caption{Total (1-Year) Ah-Throughput for Two Controls in Two Household Sizes}
\label{AhThrough}
\centering
\resizebox{0.3\columnwidth}{!}{
\begin{tabular}{r c c}
\toprule
\textbf{}                         & \textbf{1606 ft\textsuperscript{2}} & \textbf{4252 ft\textsuperscript{2}}  \\ 
\midrule
\textbf{Baseline}                 & 385 Ah              & 343 Ah               \\ 
\textbf{HEM}                     & 405 Ah              & 480 Ah               \\ 
\midrule
\textbf{Relative change} & 5.2\%               & 40\%                 \\
\bottomrule
\end{tabular}}
\end{table}

\begin{figure}[!htb]
\centering
 \includegraphics[trim=0cm 0cm 0cm 0cm, clip=true, width=0.4\textwidth]{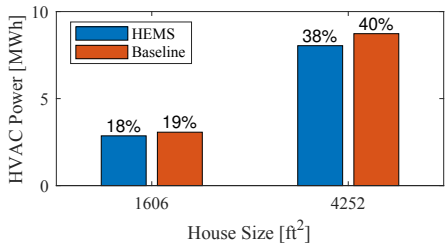}
\caption{HVAC Power and Relative Percentage of Total Power}
\label{Fig:RelativeHVACPower}
\end{figure}

It is understood that as the house size increase, the relative power required to the HVAC operation drastically increases, as shown in Figure \ref{Fig:RelativeHVACPower}.
In this figure, the percentage of the HVAC power relative to the total amount of power is denoted as the numbers over the bar plot.
In the larger house, the increased HVAC relative power will result in greater discharge events of the energy storage for larger house sizes.
Additionally, the frequency of HVAC operation is much higher for the HEM strategy compared to the baseline operation which utilizes dead-band control.Therefore, there is a correlation between increase in house size increases, and increased utilization of the energy storage. 

\begin{figure}[t]
    \centering
  \subfloat[Costs to the User \label{Fig:EffectLocation_Cost}]{%
       \includegraphics[trim=0cm 0cm 0cm 0cm, clip=true, width=0.45\textwidth]{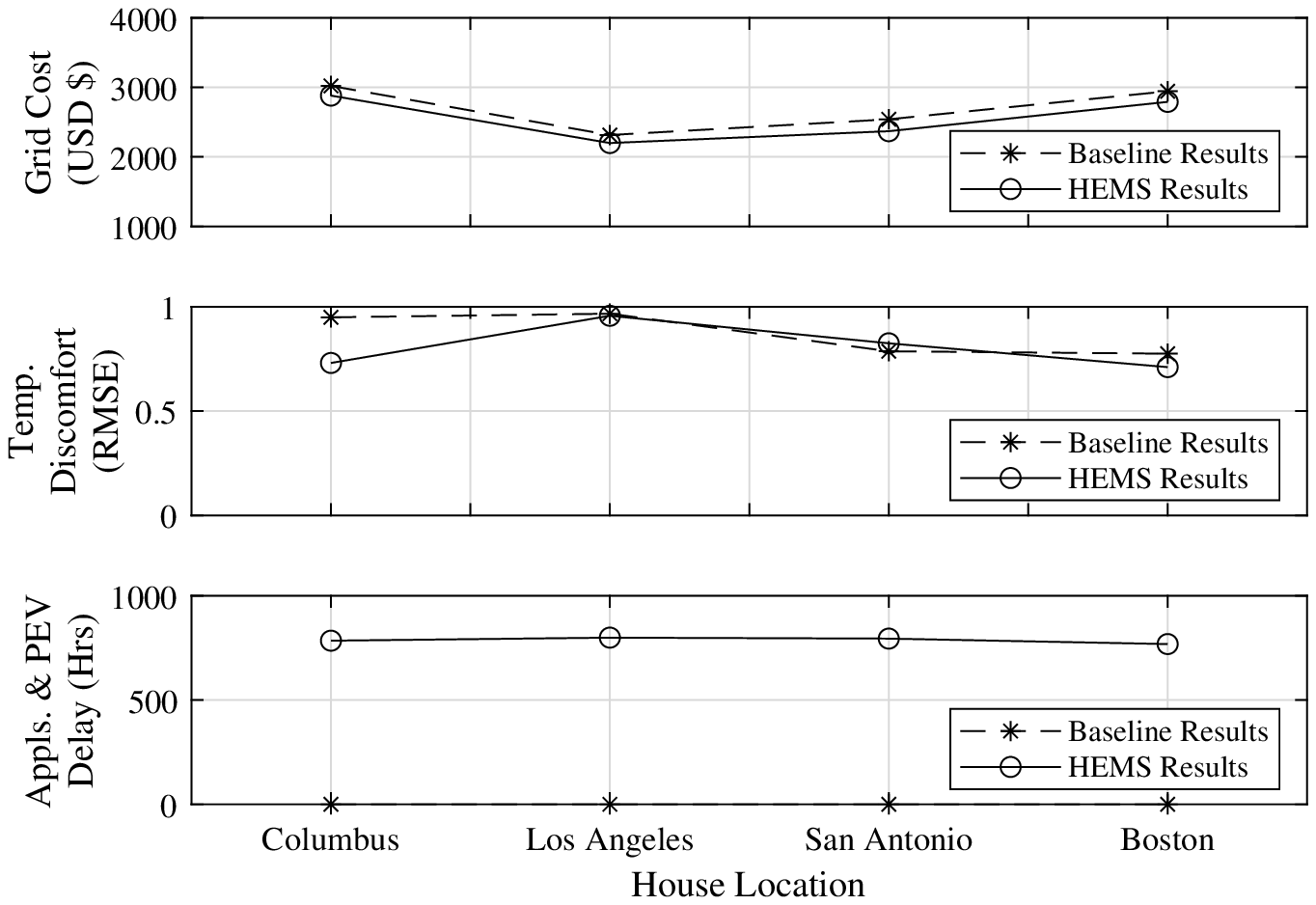}}
    \hspace{.02\linewidth}
  \subfloat[Power Deferral Efficiency \label{Fig:EffectLocation_PowerDeferral}]{%
        \includegraphics[trim=0cm 0cm 0cm 0cm, clip=true, width=0.45\textwidth]{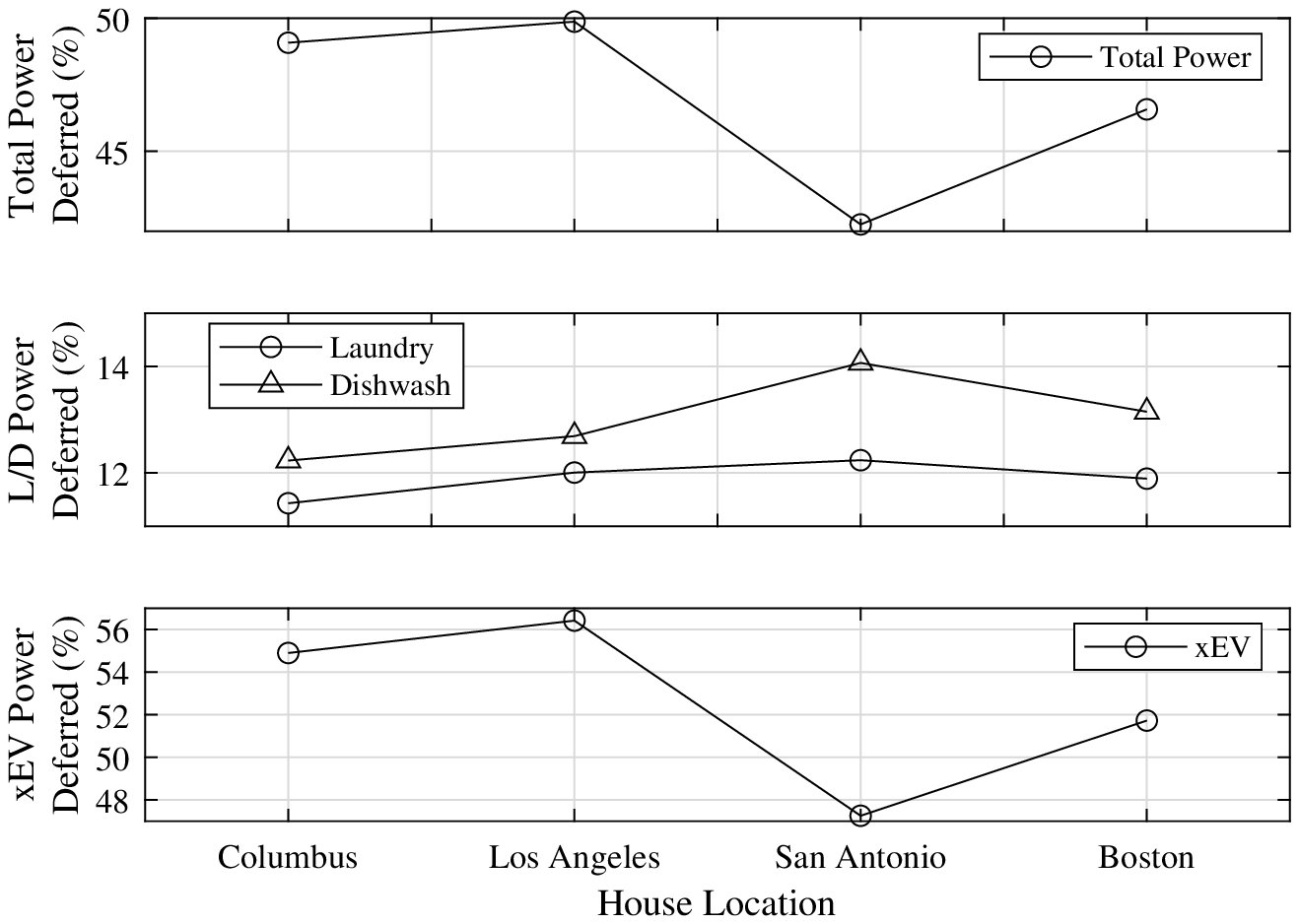}}
  \caption{Effect of household location}
\end{figure}

\subsection{Effect of House Location on Controller Performance}
The effects of house locations on both HEM strategy and baseline controller performance are analyzed by comparing cases 1, 5, 6, and 7 defined in Table \ref{Tab:CaseStudies}.

As shown in Figure \ref{Fig:EffectLocation_Cost}, for all the house locations, the average total grid cost savings of using the HEM strategy is 5.4\% compared to the baseline controller.
It is also observed that the total energy saving is almost insensitive to the house locations. On the other hand, Figure \ref{Fig:EffectLocation_Cost} shows that the HEM strategy only obtains better temperature control in Columbus compared to the other three locations, especially for Los Angeles and San Antonio.

%The HEM strategy improves the performance of HVAC compared to baseline cases when HVAC is in heating modes. 
For locations with consistent higher ambient temperatures and due to the structure of the HEMS temperature control, the resulting optimal home temperature will be close to its upper bound. This strategy decreases the overall HVAC power requirement, but leads to a higher overall temperature difference. Conversely, for locations with either lower ambient temperatures or more variability in ambient temperature over a year, the HEM strategy tends to cycle the home temperature. As a result, the largest improvement in temperature tracking compared to the baseline controller are observed in Columbus.

The deferred power metrics in Figure \ref{Fig:EffectLocation_PowerDeferral} show  that there are variations of both total and individual deferred appliance power among different locations.
This is attributed to the fact that the house sizes are different for the four locations even though the house size range is the same, as between 1500 ft\textsuperscript{2} and 2500 ft\textsuperscript{2}.
The house sizes for the four locations are listed in the first row of Table \ref{Tab:DefPowSum_Location}.
As addressed in the discussion on the effect of house size, larger house size requires greater power to operate the HVAC.
As a result, San Antonio has the lowest deferrable power efficiency, as shown in Figure \ref{Fig:EffectLocation_PowerDeferral}.
The effects of house location on the house deferral efficiency are summarized in the same figure.
%All the studied smart homes at multiple different locations are quite flexible, with a flexibility over 60\% for every case in shown. 
The deferral efficiency surpasses 42\% for every presented case, validating that the HEM strategy is effective. 
The power distribution within the four house locations is summarized in Table \ref{Tab:PowSum_Location}.

\begin{table}[!htb]
\caption{Summary of the Annual Deferrable Power Demand Depending on Location and Size of the Household.}
\label{Tab:DefPowSum_Location}
\centering
\resizebox{0.7\columnwidth}{!}{
\begin{tabular}{r | c  |  c  |  c  |  c } 
\toprule
\textbf{Location} & \textbf{Columbus} & \textbf{Los Angeles} & \textbf{San Antonio} & \textbf{Boston}  \\ 
\midrule
\textbf{Size}  & 1606 [ft\textsuperscript{2}] & 1710 [ft\textsuperscript{2}] & 2410 [ft\textsuperscript{2}] & 2183 [ft\textsuperscript{2}] \\ 
\hline \hline
Vehicle Charging [MWh] & 24.09 & 22.37 & 20.74 & 24.94 \\ 
\hline
Laundry [MWh]   & 0.49 & 0.49 & 0.49 & 0.49 \\ 
\hline
Dishwasher [MWh]  & 0.20 & 0.20 & 0.20 & 0.20  \\
\hline
Total Deferrable [MWh] & 24.77 & 24.97 & 25.85 & 29.64 \\ 
\hline
Deferred [MWh] & 12.11 & 12.47 & 10.96 & 13.78 \\  
\hline 
Deferral Efficiency [\%] & 48.89 & 49.94 & 42.40 & 46.49 \\
\bottomrule
\end{tabular}
}
\end{table}

\begin{figure}[t]
    \centering
  \subfloat[Solar Power Utilization Split \label{Fig:EffectLocation_SolarEnergyStorage}]{%
       \includegraphics[trim=0cm 0cm 0cm 0cm, clip=true, width=0.45\textwidth]{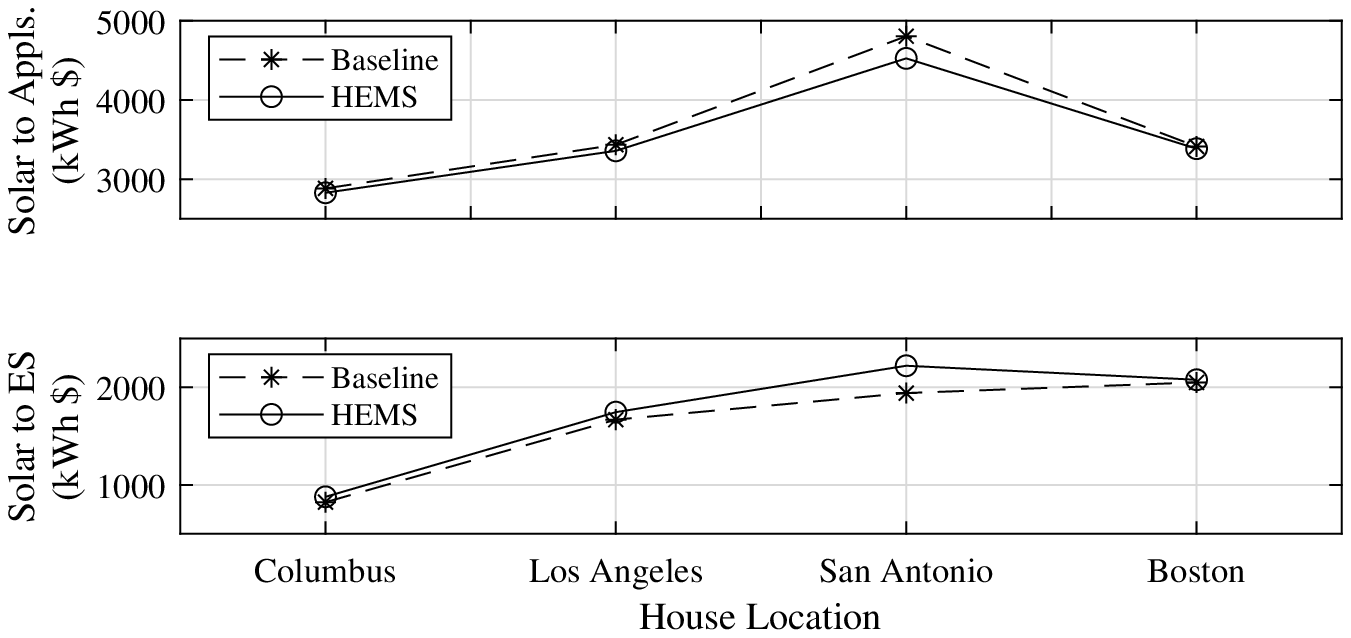}}
    \hspace{.02\linewidth}
  \subfloat[Energy Storage Behavior \label{SubFig:Location_SolarUt}]{%
        \includegraphics[trim=0cm 0cm 0cm 0cm, clip=true, width=0.45\textwidth]{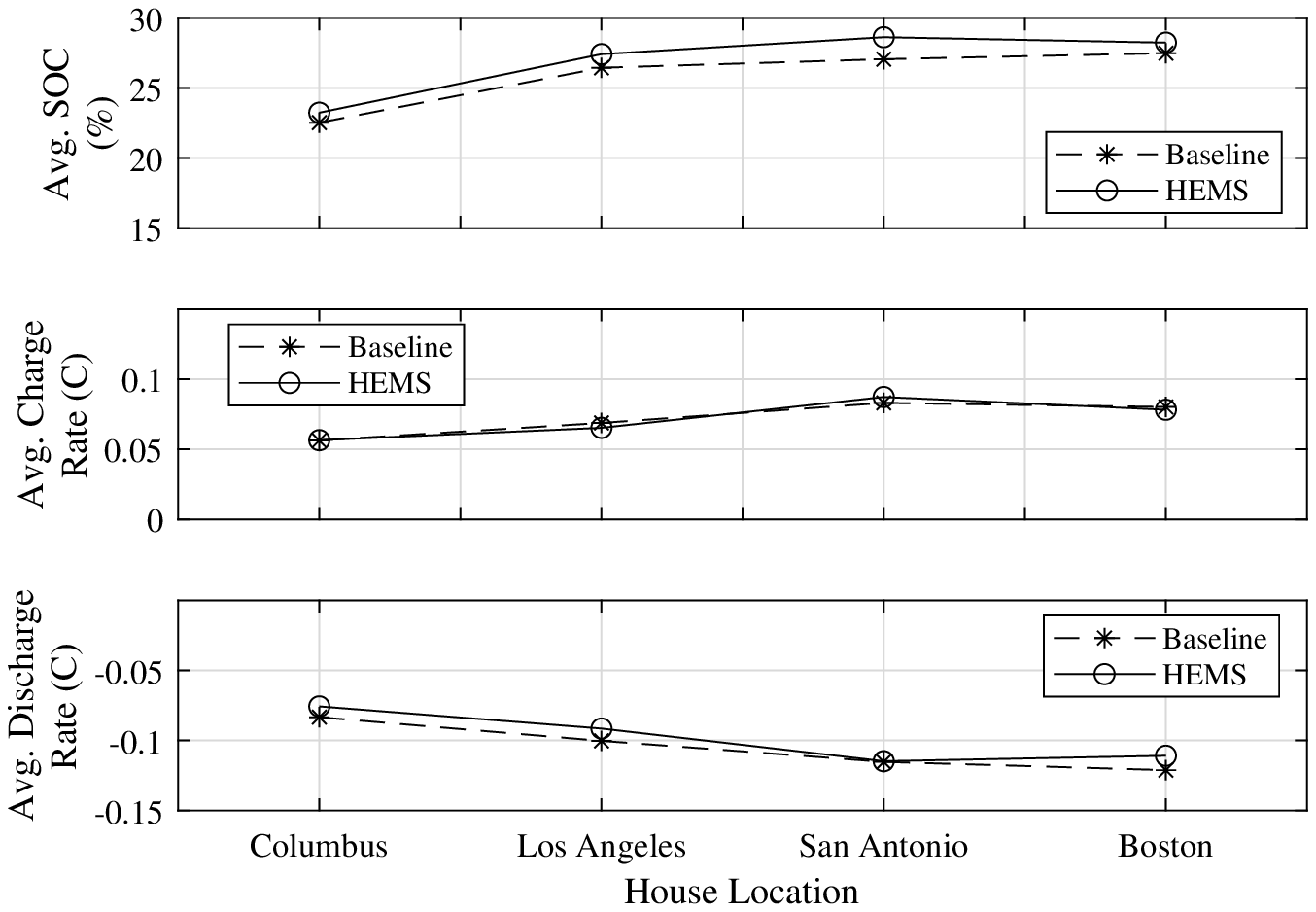}}
\caption{Solar Power Utilization and Energy Storage Depending on Location of the Household}
\end{figure}

\begin{table}[!htb]
\caption{Summary of the Annual Power Demand Depending on Location and Size of the Household.}
\label{Tab:PowSum_Location}
\centering
\resizebox{0.7\columnwidth}{!}{
\begin{tabular}{r | c  |  c  |  c  |  c } 
\toprule
\textbf{Location} & \textbf{Columbus} & \textbf{Los Angeles} & \textbf{San Antonio} & \textbf{Boston}  \\ 
\midrule
\textbf{Size}  & 1606 [ft\textsuperscript{2}] & 1710 [ft\textsuperscript{2}] & 2410 [ft\textsuperscript{2}] & 2183 [ft\textsuperscript{2}] \\ 
\hline \hline
HVAC [MWh] & 2.86 & 1.91 & 4.43 & 4.01 \\ 
\hline
Non-Controllable Act [MWh] & 7.92 & 7.85 & 7.89 & 7.93 \\ 
\hline
Total Non-Deferrable [MWh] & 10.78 & 9.76 & 12.32 & 11.94 \\ 
\hline
Total Deferrable [MWh] & 24.77 & 24.97 & 25.85 & 29.64 \\ 
\hline
Total Power [MWh] & 35.55 & 34.73 & 38.17 & 41.58 \\
\bottomrule
\end{tabular}
}
\end{table}
Finally, Figure \ref{Fig:EffectLocation_SolarEnergyStorage}
confirms that in certain cases the HEM strategy is able to utilize the solar power more efficiently by leveraging the energy storage system to lower the grid cost. 
Even if the deferral efficiency is the lowest in San Antonio, the utilization of the solar power to increase the energy storage SOC is the most significant in this location.
Hence, it is the non-deferrable power, typically HVAC, that drives
the split between storing the locally generated power or utilizing it to meet the instantaneous demand.
The utilization of the energy storage relative to the baseline is in fact much larger in San Antonio than in any other location, as shown in Figure \ref{SubFig:Location_SolarUt}.

\section{Conclusions} \label{sec:Conclusion}

% Original structure of the conclusion

This paper presents the development and comprehensive evaluation of an online HEM strategy for the coordination of loads in a smart home. Compared to a baseline strategy, the proposed HEM strategy succeeds in minimizing both grid electricity cost in response to a TOU schedule and and user discomfort by operation of deferrals.
The evaluation of the HEM performance is conducted based on the definition of systematic performance metrics, over multiple simulations with one year duration and different household characteristics. 

Results show the non-negligible impact of location, seasonality, battery size and home size on the HEM strategy performance. The HEM strategy saves on average \$160  of the electricity cost to the user per year compared to the baseline case.
The deferrable appliance are rescheduled by HEM controller to simultaneously benefit from low grid prices, renewable generation and stationary energy storage. Moreover, simulations show that the optimal strategy 
consists in charging the energy storage using the renewable sources, rather than directly using them to meet the household power demand.
Additionally, the HEM strategy is shown to provide better temperature comfort at a lower cost, compared to the traditional dead-band controller.

In all cases it is found that the energy storage is utilized more frequently by the HEM strategy, compared to the baseline strategy. However, the battery size for either the electric vehicle or the energy storage has a limited effect of the performance of the HEM strategy.
Conversely, the size and location of the household is a main factor when evaluating the controller performance. This is due to the relationship between home size and location with the HVAC operation, which is not completely deferrable and is associated to a large power demand. In fact, to meet temperature constraints, the HEM strategy is forced to operate the HVAC during high price regions. 
This is further confirmed by analyzing the deferral efficiency in locations with high HVAC requirements, such as San Antonio. In this case, the low deferral efficiency is coupled with a more aggressive utilization of the energy storage system in an attempt to delay the HVAC load to low price regions. 

Overall, the integration of renewable energy sources in the HEM strategy results in a lower consumer energy cost, independently on the ability of the controller to defer the load.

Finally, the stationary energy storage is operated more frequently by the HEM strategy compared to the baseline, which results to higher currents (C-rate) and larger spreads in SOC. Both of these conditions can lead to non-negligible battery capacity fade. Future work will consider the utilization of the battery inside the optimization problem.

\section{Acknowledgement}
The authors are grateful for the support of Ford Motor Company and the Ford-OSU Alliance Program.

\bibliography{bibfile}

\end{document}